\documentclass[aps,prd,10pt,nofootinbib,twocolumn,superscriptaddress,floatfix,notitlepage]{revtex4-1}
\pdfoutput=1

\usepackage{graphicx}
\usepackage{amsmath,amsfonts,amssymb}
\usepackage[dvipsnames]{xcolor}
\usepackage[breaklinks,colorlinks,urlcolor=MidnightBlue,citecolor=WildStrawberry,linkcolor=Fuchsia]{hyperref}
\usepackage{verbatim}
\usepackage{enumitem}
\usepackage{aas_macros}
\usepackage{subcaption}

\usepackage{soul}
\soulregister{\cite}{7}
\usepackage{xcolor}
\usepackage[table]{xcolor}

\definecolor{pAbove}{HTML}{D9EAF7} 
\definecolor{pEqual}{HTML}{FFF2CC} 
\definecolor{pBelow}{HTML}{FADBD8} 

\usepackage{graphicx}
\usepackage{amsmath,amsfonts,amssymb}
\usepackage[dvipsnames]{xcolor}
\usepackage[breaklinks,colorlinks,urlcolor=MidnightBlue,citecolor=WildStrawberry,linkcolor=Fuchsia]{hyperref}
\usepackage{verbatim}
\usepackage{enumitem}
\usepackage{aas_macros}
\usepackage{subcaption}

\newcommand{\td}{{\rm d}}

\newcommand{\be}{\begin{equation}}
\newcommand{\ee}{\end{equation}}
\newcommand{\bea}{\begin{equation} \begin{aligned}}
\newcommand{\eea}{\end{aligned} \end{equation}}

\def\lsim{\mathrel{\raise.3ex\hbox{$<$\kern-.75em\lower1ex\hbox{$\sim$}}}}
\def\gsim{\mathrel{\raise.3ex\hbox{$>$\kern-.75em\lower1ex\hbox{$\sim$}}}}

\usepackage[capitalise]{cleveref}
\usepackage[activate={true,nocompatibility},final,kerning=true,factor=1100,stretch=10,shrink=10]{microtype}
\usepackage{academicons}
\usepackage{fontawesome5}
\definecolor{orcidlogocol}{rgb}{0.65, 0.807, 0.223}
\newcommand{\orcid}[1]{$\,$\href{https://orcid.org/#1}{\textcolor{orcidlogocol}{\faOrcid}}}

\begin{document}

\title{Phenomenological implications of cosmic (super)strings with time-varying tension}

\author{Anastasios Avgoustidis\orcid{0000-0001-7247-5652}}\email{anastasios.avgoustidis@nottingham.ac.uk}

\author
{Edmund J. Copeland\orcid{0000-0003-3959-6051}}\email{edmund.copeland@nottingham.ac.uk}

\author{Adam Moss\orcid{0000-0002-7245-7670}}\email{adam.moss@nottingham.ac.uk}

\author{Juhan Raidal\orcid{0009-0002-2600-3632}}
\email{juhan.raidal@nottingham.ac.uk}
\affiliation{School of Physics and Astronomy, The University of Nottingham, Nottingham, NG7 2RD, UK}

\author{Filippo Revello\orcid{0000-0002-1887-8351}}
\email{filippo.revello@kuleuven.be}
\affiliation{Instituut voor Theoretische Fysica, KU Leuven, Celestijnenlaan 200D, B-3001 Leuven, Belgium}
\affiliation{Leuven Gravity Institute, KU Leuven, Celestijnenlaan 200D, B-2415 Leuven, Belgium}

\author{Gonzalo Villa\orcid{0000-0002-0132-1178}}
\email{gv297@cam.ac.uk}
\affiliation{DAMTP, Centre for Mathematical Sciences, University of Cambridge,
Wilberforce Road,Cambridge, CB3 0WA, UK}

\begin{abstract}

Cosmic strings and cosmic superstrings could have formed in the early universe. In a particular class of models, the tension associated with these defects is time dependent, either because the strings themselves are composed of evolving moduli fields, or because they are directly coupled to evolving scalar fields leading to an effective time dependence in the string tension. Here we take seriously the possibility of this time dependence and investigate the phenomenological consequences of such an evolution, comparing the results to the more traditional case of a fixed time independent tension. In particular, for five different time varying models, motivated by different possible forms of evolution of the string tension, we evaluate their Gravitational Wave (GW) contributions as well as their impact on the Cosmic Microwave Background (CMB). Performing an MCMC analysis of the models we find there are indeed interesting spectral signatures in the GW emission at frequencies between the current PTA and interferometer data, but the signatures are not unique to these strings, making them hard to pin down using current data. However, we do find cosmic strings with evolving tension fit the GW data as well as other favoured models.

\end{abstract}

\maketitle

\section{Introduction}
Cosmic strings are among the few well-motivated remnants of high energy physics which could still be found in the late universe.
These one-dimensional filaments of energy are generated as topological defects in quantum field theory~\cite{Kibble:1976sj} (which are in particular generic in supersymmetric grand unified theories~\cite{Jeannerot:2003qv}) or as fundamental objects in string theory~\cite{Polchinski:2004ia}.

In addition to their intrinsic interest as windows into the high energy world, cosmic strings are potentially relevant for cosmology.
Their dynamics lead to the formation of networks of strings~\cite{Kibble:1984hp} that form a small,  constant fraction of the total energy of the Universe, in a so-called scaling regime. In the early days of cosmic string research, 
the perturbations they sourced were thought to be a candidate for structure formation, rivalling the inflationary universe scenario, but once the multiple peak structure in the CMB was discovered, they were quickly ruled out as such candidates for structure formation~\cite{Albrecht:1997nt}.
However, that doesn't rule out their formation in the Early Universe. For example the post-inflationary model for the QCD axion as dark matter has these topological defects as the sources of the dark matter (see e.g.~\cite{Marsh:2015xka}).
The CMB cannot constrain these because they become unstable around the QCD phase transition.

Moreover, the scaling regime is believed to predict\footnote{The primary decay route for cosmic strings, whether it be through particle creation or GWs remains one of the ongoing questions in the subject - see for example~\cite{Hindmarsh:2021mnl} for the case of particle production. In this paper, we assume they decay primarily into GWs.} a stochastic GW background with a non-negligible energy density spread across many orders in magnitude in frequency.
Within the frequency regime probed by current and future detectors, its shape is sensitive in particular to the equation of state of the dominant energy components present in the early universe~\cite{Cui:2017ufi}.
A detection of the GW spectrum of cosmic strings would thus provide a spectacular window into pre-Big Bang Nucleosynthesis physics.
Along these lines, recent results by Pulsar Timing Arrays (PTAs) \cite{NANOGrav:2023hvm,EPTA:2023fyk,Reardon:2023gzh,Xu:2023wog} have reported evidence for a gravitational wave background with frequencies around $10^{-8}$ Hz.
While the astrophysical explanation arising from  mergers of supermassive black hole binaries remains the most likely source of the signal, it cannot be accommodated by the current simplest modelling~\cite{Ellis:2023dgf}.
Given the frequency range, it is not surprising the cosmological signals have also been studied in this context, with ``cosmic superstrings'' providing a good fit to the data~\cite{Ellis:2023oxs,Ellis:2023tsl,Avgoustidis:2025svu}.

One thing that makes cosmic strings exciting is that the nature of the scaling regimes for stable cosmic string networks are therefore probed by current and future experimental efforts, with the string tension $\mu$ in Planck units the most constrained quantity, as it sets the fraction of energy density in strings with respect to that of the universe during scaling.
The CMB scalar anisotropies now provide a model-independent bound whilst PTAs give tighter but model-dependent constraints on the string tension - for recent details see \cite{Raidal:2026cpb}. However, all these constraints assume a time independent string tension throughout cosmic history. Whilst this is a standard assumption, it may not be the case in reality. Of the rich dynamics giving rise to cosmic string physics, much of it leads to the formation of strings with a time-dependent tension. For example, in gauge theory, if the string forming symmetry breaking has a temperature-dependent order parameter, the tension redshifts with temperature until the true minimum of the effective potential is reached~\cite{Emond:2021vts}.
More generally, when the fields undergoing the phase transition are coupled to others, including those responsible for the formation of the cosmic strings, the tension of the string depends on the overall dynamics of the system.
Cosmic strings with varying tension have been studied recently in the context of string theory~\cite{Conlon:2024uob,Revello:2024gwa}, with applications~\cite{Ghoshal:2025tlk,Brunelli:2025eif,SanchezGonzalez:2025uco,Conlon:2025mqt,Brunelli:2025ems,Brunelli:2026ndh,Brunelli:2026qkp} usually focusing on early universe dynamics.

Dynamical scalar fields in the late universe have been proposed as candidates for tackling fundamental physics questions, such as the cosmological constant problem~(see~\cite{Kaloper:2013zca,Kaloper:2014fca,Burgess:2021obw,Burgess:2025vxs} for some recent attempts, and~\cite{Padilla:2015aaa} for a review), or observational issues like the Hubble tension~\cite{Poulin:2018cxd,Copeland:2023zqz,Smith:2025uaq} - for a review see \cite{Kamionkowski:2022pkx}.
As per the above, these may induce variations in the tension of cosmic strings.
In this paper we take this possibility seriously and explore the sensitivity of the CMB and GW detectors to cosmic strings with varying tension.

The rest of the article is organised as follows.
In section~\ref{sec:vtension} we describe cosmic strings with varying tension in the context of string theory and discuss the backreaction of the strings on the dynamics of the fields which are responsible for the variation of the tension, finding a negligible effect, thereby allowing us to evolve the fields without any backreaction.
The rest of the paper is independent of that section, and concerns more phenomenological issues associated with time dependent tensions.
In section~\ref{sec:modelling} we discuss the impact of the time varying tensions on the modelling of the networks.
Section~\ref{sec:GWs} reviews the GW spectrum of such a cosmic string network, which we further explore in section~\ref{sec:tension_models} by concentrating on five evolving classes of phenomenological models describing the evolution of the tension. We then discuss the effect of these networks on the CMB anisotropies in section~\ref{sec:CMB}, while in section~\ref{sec:results} we present and discuss our results. Finally, we conclude in section~\ref{sec:conclusions} and have two appendices at the end.

\section{Cosmic (super)strings with varying tension}\label{sec:vtension}

This section describes the string theory motivation behind cosmic strings with varying tension.
We discuss known and some new examples of how this phenomenon arises, and discuss its cosmological relevance as a consequence of the generic exponential potentials in string compactifications.
String tensions are always field-dependent in string theory models, where dimensionless quantities are determined dynamically by the expectation values of fundamental fields.
Particularly attractive for present purposes are moduli fields which, in cosmological backgrounds, are dynamical for several e-folds, inducing a long-lived variation of the string tension.\footnote{\label{ft:stabilisation}While moduli like the string dilaton or the volume modulus must be stabilised before Big Bang Nucleosynthesis as to accommodate observational bounds, other moduli need not induce variation of fundamental constants in the Standard Model.
Although still constrained by e.g. CMB (see~\cite{Agarwal:2026dko} for a recent discussion in the context of moduli) and fifth force experiments (see~\cite{Joyce:2014kja,Burrage:2017qrf} for reviews), these could be dynamical after reheating.
}
The reason behind such cosmologically relevant dynamics are generic approximate scale invariances of extra dimensional theories~\cite{Burgess:2020qsc}, which result in exponential potentials.
Such potentials are well known to lead to scaling solutions where the field tracks the energy density of the Universe~\cite{Wetterich:1987fm,Copeland:1997et,Ferreira:1997hj}.

\subsubsection{Moduli-dependent tension}

In a string theoretic setting, the dimensionless ratio between the tension of both fundamental and field-theory strings and the Planck scale depends, as any other dimensionless quantity, on the expectation values of the underlying fields.
As an example, the fundamental string tension in the 4d Einstein frame is given by the expression
\begin{equation}
    \mu \equiv m_s^2= \frac{g_s^{1/2}M_P^2}{\mathcal{V}},
\end{equation}
where $\mathcal{V}$ is the volume of the extra dimensions (in string units), $g_s$ the string coupling, determined by the vacuum expectation value of the dilaton, and $M_P$ the 4-dimensional Planck mass. For a bound state of $p$ F$1$ and $q$ D$1$ strings, also known as a $(p,q)$-string, the analogous formula is
\begin{equation}\label{eq:pqstring}
     \mu_{(p,q)}
    =
    \frac{M_P^2}{\mathcal V}
    \sqrt{
        g_s p^2 + \frac{q^2}{g_s}
    } .
\end{equation}
More generally, one may also obtain a string-like object in 4d by wrapping a D$p-$brane along an internal $(p-1)$-cycle $\Sigma_{p-1}$. Its tension equals that of the original brane, multiplied by the (string frame) volume of $\Sigma_{p-1}$ in string units:
\begin{equation}
    \mu_{\mathrm{D}p}
    = \frac{g_s^{-1/2} M_P^2}{\mathcal V}\,
      \frac{\mathrm{Vol}(\Sigma_{p-1})}{( 2 \pi \sqrt{\alpha'})^{p-1}}\,
\end{equation}
where $2 \pi \alpha'$ is the inverse of the fundamental string tension. The tension is therefore a product of power laws of geometric quantities such as cycle volumes, as well as the dilaton. Since the corresponding canonically normalised moduli, $\phi_i$, are logarithmic functions of these quantities,
\begin{equation}
\phi_i = c_i M_P \log \left(\frac{X_i}{X_{i,0}}\right),
\end{equation}
where \(X_i\) denotes the relevant geometric quantity and \(c_i\) is fixed by the corresponding moduli-space metric, any power-law dependence on \(X_i\) becomes exponential when expressed in terms of the canonically normalised fields. The tension can therefore be written schematically as
\begin{equation}
\mu(\phi_i) = \mu_0 e^{\alpha_j \phi_j/M_P},
\end{equation}
where Einstein summation over repeated moduli indices is understood.  For the universal volume directions, typically $\alpha_j <0$, so that a decreasing string tension is the most natural scenario (both in field space and, as we will see, in time). On the other hand, more complicated geometries such as a Calabi-Yau with a fibred structure can also exhibit a growing tension. In the context of type IIB string theory, phenomenologically interesting examples - such as D$3$ branes wrapping 2-cycles and NS$5$ branes wrapping 4-cycles - have been studied in \cite{Brunelli:2025ems,Brunelli:2025eif,Brunelli:2026qkp,Brunelli:2026ndh}. 
 
 Another option, which has so far attracted less attention in the literature, concerns the case where the tension also inherits a dependence on axions from the DBI action. The cleanest example arises from a $(p,q)$ string in the presence of a non-trivial profile for the RR axion
$C_0$ \cite{Schwarz:1995dk}. In that case, Eq. \eqref{eq:pqstring} is modified to
\begin{equation}\label{eq:tax}
    \mu_{(p,q)}
    =
    \frac{M_P^2}{\mathcal V}
    \sqrt{
        g_s\left(p-qC_0\right)^2
        + \frac{q^2}{g_s}
    } .
\end{equation}
As a more realistic case, one may consider a D$3$ brane wrapping a $2$-cycle $\Sigma_2$ and with worldvolume flux. Using dimensionless quantities in string units, the $2-$cycle size modulus and its associated $B_2$ axion can be defined as
\begin{equation}\label{eq:axdef}
v=\frac{{\rm Vol}(\Sigma_2)}{ (2 \pi)^2\alpha'},
\qquad
b=\frac{1}{(2 \pi)^2 \alpha'}\int_{\Sigma_2} B_2.
\end{equation}
Massless excitations of open strings ending on the brane give rise to an additional abelian gauge field $A_1$, which can thread a (quantised) worldvolume flux $F_2= \rm{d} A_1$ along the same 2-cycle
\begin{equation}
    \frac{1}{2\pi}\int_{\Sigma_2} F_2 = f_2.
    \qquad \quad f_2\in\mathbb{Z}.
\end{equation} 
Thus, its four-dimensional Einstein-frame tension takes schematically the form
\begin{equation}\label{eq:D3w}
\mu_{\rm D3}
=
\frac{-g_s^{1/2}M_P^2}{\mathcal V}
\sqrt{v^2+(b+f_2)^2}.
\end{equation}
The form of Eq.~\eqref{eq:D3w} is reminiscent of the expression for the 4d scalar potential in models of axion monodromy inflation \cite{Silverstein:2008sg,McAllister:2008hb}. This is because the latter arises from D$5$-branes wrapping an internal 2-cycle in the compactification manifold, which are however spacetime filling in 4d. In both cases, the expression for the tension (potential) becomes linear in $b$ for $b\gg v$ on a fixed $f_2$ branch.\footnote{The axion shift symmetry survives only through the monodromy $b\rightarrow b+1$ and $f_2\rightarrow f_2-1$.} In the context of axion monodromy inflation, this regime of large field excursion is thought to suffer from potentially dangerous backreaction on the stabilised value of the scalar moduli (such as $v$) \cite{Hebecker:2014kva,Baume:2016psm}. However, the backreaction on the bulk stabilised moduli is expected to become less severe in the case of strings arising from wrapped branes. Unlike axion monodromy inflation, the wrapped D$3$-branes are not spacetime filling, and their energy density contributes to the 4d scalar potential only at the location of the strings.

Let us end with a cautionary remark: despite this theoretical variety of possible scenarios, their viability depends crucially on the details of moduli stabilisation. For instance, in standard type IIB flux compactifications the \(C_0\) axion is typically stabilised together with the dilaton (and complex-structure moduli) already at tree level, making it generically an unsuitable candidate for time-dependent tensions \cite{Giddings:2001yu}.

\subsubsection{Scalar dynamics} \label{sec:scalar-dyn}

As for string tensions, individual contributions to scalar potentials in string compactifications typically depend on geometric quantities through power laws. The main difference is that the scalar potential generally receives several such contributions. After expressing the geometric quantities in terms of canonically normalised moduli, the potential takes the form of a sum of exponentials of (or linear combinations of) various moduli. For simplicity, let us consider the more straightforward case of a single scalar field $\phi$ with an exponential potential
\begin{equation}\label{eq:scalar-pot}
    V=V_i e^{-\lambda(\phi-\phi_i)/M_p} \, ,
\end{equation}
assumed to be the dominant contribution in a certain regime of parameter space, and where $\lambda$ and $\phi_i$ are constants. In an FLRW background, its equation of motion reads
\begin{equation}
    \ddot{\phi}+3H\dot{\phi}+\frac{\partial V}{\partial \phi}=0 \, ,
\end{equation}
where $H = \dot{a}/a$ is the physical Hubble rate. For potentials of the form dictated by Eq.~\eqref{eq:scalar-pot} it is straightforward to check that there is a solution of the form
\begin{equation}\label{eq:scaling-sol}
    \phi=\phi_i+\frac{2M_p}{\lambda}\log\left(\frac{t}{t_i}\right)\, ,
\end{equation}
for an appropriate value of $t_i$ and $\phi_i$ for a power-law scale factor $a(t) \sim t^{2/n}$.\footnote{Note that the dependence on $n$ is implicit through either $t_i$ or the coefficient $V_i$.} A dynamical system analysis \cite{Wetterich:1987fm,Copeland:1997et,Ferreira:1997hj} shows that this solution is an attractor provided that $\lambda^2 > n$. An important property of this solution is that the kinetic and potential energies of the field are a meaningful constant fraction of the total energy density of the Universe, which is usually called a tracker regime. Similarly, there exist more complicated tracker regimes where axions that couple to moduli through their kinetic term can also co-dominate the evolution \cite{Cicoli:2020cfj,Cicoli:2020noz,Revello:2023hro,Licciardello:2025fhx}. Notice that axions defined as in Eq.~\eqref{eq:axdef} are not canonically normalised, and along a tracker they have the same time-dependence as the non-canonically normalised cycle (the cycle size). For example, in the D3 case, assume that $v$ parametrises an overall
two-cycle scale, with fixed $g_s$ and internal shape, so that
$\mathcal{V}\propto v^3$. Keeping only $v$ and the associated $B_2$
axion $b$, the bulk kinetic terms take the schematic form
\begin{equation}
    \mathcal{L}_{\rm kin}
=
-\frac{M_P^2}{2v^2}
\left[
6(\partial v)^2+c_b(\partial b)^2
\right],
\end{equation}
where $c_b>0$ is compactification dependent. With the potential given by Eq.~\eqref{eq:scalar-pot}, the system admits a stable, modulus-axion tracker if $\lambda^2+\sqrt{\frac{2}{3}}\lambda>6$. It is characterised by
\begin{equation}\label{eq:modaxtr}
v(t)\propto e^{\frac{\phi}{\sqrt{6}M_P}} \propto t^{2/(\sqrt{6}\lambda)},
\qquad
b(t)=b_i+\kappa\,v(t),
\end{equation}
with $\kappa$ a constant and $b_i$ the initial condition for $b$.

Another possibility, especially relevant for steep potentials, arises when the kinetic energy of a single field comes to dominate over all other sources (matter and/or scalar potential), giving rise to what is known as kination. A single, kinating scalar evolves as
\begin{equation}\label{eq:kination-sol}
    \varphi(t)=\varphi_i + \sqrt{\frac{2}{3}}\,M_p
    \log\left(\frac{t}{t_i}\right),
\end{equation}
where $\varphi=\phi$ or $f_b b$, with $f_b$ the axion decay constant, hence $\varphi$ can either be a modulus or an axion.\footnote{In the latter case, the moduli determining its canonical normalisation (\emph{i.e.} the decay constant $f_b$) must be frozen.} Eq \eqref{eq:kination-sol} implies a stiff equation of state, $w_\phi\simeq 1$ and an energy density in $\phi$ decreasing rapidly with the scale factor, $\rho_\phi\propto a^{-6}$. While kination is typically unstable, it can provide an additional transient phase of rapid scalar evolution before a tracker phase, which is well motivated in the context of string compactifications \cite{Gouttenoire:2019kij,Conlon:2022pnx,Apers:2022cyl,Apers:2024ffe}.

In a realistic scenario, the kination and tracker regimes are usually a transient between two phases where the expectation value of the field is constant.
At early times, for a light scalar field at some initial position $\phi_0$, the field will remain frozen due to Hubble friction until the Hubble scale is of order of its mass, $H^2\sim \frac{\partial^2 V}{\partial \phi^2}|_{\phi_0}$, whereby after a short period it will enter the tracker regime.
In addition, the scalar potential is expected to have a minimum in order to stabilise the field (see however footnote~\ref{ft:stabilisation}), and when the field reaches its minimum it starts oscillating, behaving like a pressureless contribution to the energy budget of the Universe.
At this point, the expectation value of the field averages to a constant on cosmological timescales. The rolling phase, whether characterised by tracking or kination, corresponds to the period where the field interpolates between its initial and final values.

A natural implication of Eq.~\eqref{eq:scaling-sol}, combined with the exponential dependence of the string tensions on the moduli, is that they suggest that moduli tracker and kination epochs will be characterised by a power-law time-dependence of the string tensions, namely 
\begin{equation} \label{string-tension-1}
    \mu(t) = \mu_i \left(\frac{t}{t_i}\right)^{-m},
\end{equation}
for suitable values of the constants $m,\,t_i$ and $\mu_i$. From string-theoretic considerations, $m$ is expected to be $\mathcal{O}(1)$, and more precise bounds can be derived under suitable assumptions \cite{Brunelli:2026ndh}. Moreover, while for the vanilla case of a single scalar rolling down an exponential potential one typically expects $m \geq 0$, more involved cases with multiple scalars and/or shrinking internal cycles can also give rise to $m<0$ \cite{Brunelli:2026ndh}.  Further, it need not be a simple power law dependence. For example, for 
the tension dependence on the moduli and axions in Eqs~\eqref{eq:tax} and~\eqref{eq:D3w} they can give rise to more complicated behaviour. A kinating axion (see \emph{e.g.} \cite{Gouttenoire:2019kij}),  evolves according to Eq.~\eqref{eq:kination-sol} and thus can give rise to a tension of the form
\begin{equation}\label{eq:logaxion}
    \mu(t) = \mu_i \sqrt{1+ \left[\log \left( t/t_i\right)\right]^2} \simeq \mu_i \log \left( t/t_i\right)
\end{equation}
for values of $t \gg t_i$, where again $\mu_i$ and $t_i$ denote initial conditions (different from the ones of the previous equation).
For more general tracker solutions, the modulus and axion dependence can compete with each other and give rise to tensions that are not monotonic functions of time. As an example, consider an axion-modulus tracker of the form Eq.~\eqref{eq:modaxtr}, assuming fixed $g_s$ and $\mathcal{V}\propto v^3$. Choosing $t_i$ such that $b(t_i)+f_2=0$, the wrapped D3-brane tension Eq.~\eqref{eq:D3w} evolves as
\begin{equation}
    \frac{\mu(t)}{\mu_i}
=
\left(\frac{t}{t_i}\right)^{-2r}
\sqrt{
1+\kappa^2
\left[
1-\left(\frac{t}{t_i}\right)^{-r}
\right]^2
},
\quad 
r=\frac{2}{\sqrt{6}\lambda}.
\end{equation}
This competition can generate a dip followed by a local bump in the
tension, provided $\lambda^2+\sqrt{\frac{2}{3}}\lambda-6>24c_b$.\\
To summarise, it is quite clear that there exists a plethora of possibilities for the time dependence of the string tension. While we have illustrated this explicitly along cosmological attractors, such as tracker solutions, the range of possibilities becomes even broader once we allow for transients in the evolution of the fields themselves. This justifies the phenomenological approach adopted in the remainder of the paper, where we consider several representative forms of time-varying tension without committing to a specific microscopic realisation. In particular, we will study simple models interpolating between asymptotic regimes of constant or power-law tension, as well as more abrupt features such as steps or transient spikes.

\subsubsection{String network backreaction}\label{sec:bacreaction}

Let us now consider a perturbation to the system above due to a network of $N$ non-interacting strings whose tension depends on the field $\phi$.
The action reads
\begin{equation}
\begin{split}
    S=& \int{d^4x}\sqrt{-g} \left(\frac{M_p^2}{2}R-\frac{1}{2}g^{\mu \nu}\nabla_\mu \phi \nabla_\nu \phi -V(\phi)\right)\\
    & -\sum_{i=1}^N\int{d^2\sigma \sqrt{-h(X_i(\sigma))}\mu(\phi(X_i(\sigma)))}\, .
\end{split}
\end{equation}
where $h_{ab}(X_i(\sigma))$ is the pullback of the ambient metric onto the worldsheet of the string, located at coordinates $X_i$.

The equation of motion of $\phi$ now acquires new localised source terms due to the backreaction of the strings on the field:
\begin{equation}\label{eq:eom}
\begin{split}
    \ddot{\phi}&+3H\dot{\phi}+\frac{\partial V}{\partial \phi}-\frac{\nabla^2 \phi}{a^2} \\ &+\sum_{i=1}^N \int{d^2\sigma\sqrt{-h}\frac{\partial \mu}{\partial \phi }\delta^{(4)}(x-X_i(\sigma))}=0 \, .
    \end{split}
\end{equation}
For the particular system we have in mind, the tension features an exponential dependence on the field, $\mu=\mu_0e^{-A(\phi-\phi_0)/M_p}$,  for some constant $A$, so the source term is proportional to the total energy of the strings.
Let us decompose the field into a background, spatially homogeneous term $\bar{\phi}(t)$ plus a spacetime-dependent perturbation $\delta \phi(x,t)$.
Averaging the equation of motion over space gives, at leading order in the perturbation,
\begin{equation}
    \ddot{\bar{\phi}}+3H\dot{\bar{\phi}}+\frac{\partial V}{\partial \phi}-A \frac{\rho_s}{M_p} =0\, ,
\end{equation}
where $\rho_s$ is the average energy density contained by the strings.\footnote{We have defined $\rho_s=\frac{1}{\rm{V}}\int{d^2\sigma \sqrt{-h}\int d^3x\sqrt{-g} \delta^{(4)}(x-X_i(\sigma))}\mu(\sigma)$ which, in a gauge $X_i^{(0)}=t$, reproduces $\rho_s=\mu \sum_{i=1}^N L_i/\rm{V}$, with $L$ and $\rm{V}$ appropriately regulated string spacelike length and volume, respectively.}
Backreaction of the string network on the scaling solution above is thus negligible if, throughout cosmic evolution,
\begin{equation}
    \left|\frac{A \frac{\rho_s}{M_p}}{\frac{\partial V}{\partial \phi}} \right| \ll 1 \, .
\end{equation}
For an exponential potential, $\partial V/ \partial \phi = - \lambda V/M_p$ so, for order one $A/\lambda$ (which is the case in string compactifications), the question is whether the energy density in cosmic strings is comparable to that in the scalar field's scalar potential.
For cosmic strings in a scaling solution, in terms of the total energy density $\rho$, their fractional energy density $\Omega_s=\rho_s/\rho\sim G\mu$, with $8 \pi G=1/M_p^2$ being Newton's constant.
In the perturbative regime $G\mu \ll 1$ this is much smaller than $V/\rho \sim 1/n$ in the scaling solution above, so we conclude that string networks close to scaling do not modify the scaling solution in Eq.~\eqref{eq:scaling-sol} qualitatively.
\footnote{Other scenarios involving cosmic strings with varying tension study the case where a distribution of small loops exists in the Universe.
The relative energy density in string loops grows in this case, and eventually becomes the dominant contribution to the energy density of the Universe, which slows down the evolution of the field~\cite{SanchezGonzalez:2025uco,Brunelli:2025eif,Conlon:2025mqt,Brunelli:2026ndh}.}

Another question related to backreaction concerns the validity of perturbation theory in the discussion above, which assumes $\delta \phi(x,t)\ll \bar{\phi}$.
The main issue is that the string is a codimension two object, which locally sources a logarithmic profile for $\phi$.
For superhorizon strings as those discussed here this could be an issue, but this is not the case because the coupling between the string and the field has a Planckian strength suppression.
In more detail, the field profile sourced by an infinitely long, straight string in Minkowski spacetime is, at a distance $r$ from the string, 
\begin{equation}
    \phi=\phi_0-\frac{A \mu}{2 \pi M_p}\log(r/r_0)\, ,
\end{equation}
as follows from Eq.~\eqref{eq:eom} when specialized to a field without potential in Minkowski spacetime. 
Here $\phi_0$ would be determined by a UV calculation (e.g. closed strings scattering off the long string would determine the tension at the string scale) and the evolution of $\phi$ is computed radially outwards.
Expressing the fields in Planck units, we observe that the maximum displacement of $\phi$ from its UV value is of order $G\mu \log(r_{\text{max}}/r_0)$, which is negligible compared to the $\bar{\phi}/M_p \sim\mathcal{O}(1)$ studied here for reasonable values of $r_{\text{max}}$.\footnote{The $G\mu$ suppression we observe cannot be overcome even if $r_0=1/M_p$ and $r_{\rm max}=1/H_0\simeq 10^{60}/M_p$}
(In practice, for string networks, $r_{\rm max}$ is set in order of magnitude by the typical interstring separation, which is also the typical curvature radius of the strings.
This is parametrically similar to the Hubble scale, which is also of the same order of magnitude as the mass of the field, which also cuts off the radial evolution.)

We close this section by stressing here that the absence of backreaction is a consequence of there being two scales in the problem: the characteristic string tension $\mu$ and the Planck scale $M_p$ which sets the coupling of the string to the scalar under consideration.
Backreaction would be important in the case where both scales are parametrically similar, since then the ratio $(\phi(r_1)-\phi(r_0))/(\phi(r_1)+\phi(r_0))$ would be of order one and there would be no clean separation between the background and spatial profile of the field.

\section{String network modelling}\label{sec:modelling}

Following the discussion in sections~\ref{sec:scalar-dyn} and \ref{sec:bacreaction}, the local dynamics of the string can be calculated evaluating $\phi$ at its background value, given by Eq.~\eqref{eq:scaling-sol}.
Recalling Eq.~\eqref{string-tension-1} we thus parametrise the time-varying string tension as
\be\label{eq:tension_power}
G\mu(t) = G\mu_0 \left(\frac{t}{t_0}\right)^{-m}, \qquad m \in \mathbb{R},
\ee
where $t$ is physical time, $t_0$ is the age of the universe today, and $G\mu_0 = G\mu(t_0)$ is the tension today.

In physical time the velocity-dependent one-scale (VOS) equations governing the dynamics of a cosmic string network with time varying tension are given by \cite{Revello:2024gwa}
\be\label{eq:VOSL}
\frac{dL}{dt} = HL(1+v^2) + \frac{\tilde{c}}{2}\,v + \frac{L}{2\mu}\frac{d\mu}{dt}\,v^2,
\ee
\be\label{eq:VOSv}
\frac{dv}{dt} = (1-v^2)\left[\frac{k(v)}{L} - 2Hv - \frac{1}{\mu}\frac{d\mu}{dt}\,v\right],
\ee
where $\tilde{c}$ is the loop-chopping efficiency, $L$ is the correlation length, $v$ is the RMS velocity and
\be
k(v) = \frac{2\sqrt{2}}{\pi}\,\frac{(1-8v^6)(1-v^2)(1+2\sqrt{2}\,v^3)}{1+8v^6}.
\ee
Note that the functional form of the phenomenological parameter $k(v)$ is unchanged from the constant-tension case \cite{Revello:2024gwa}.

It is convenient to rewrite the VOS equations in terms of $\xi = L/t$, which gives
\be \label{eq:VOS1}
\frac{d\xi}{dt} = \frac{\xi}{t}\left[Ht(1+v^2) - 1 + \frac{\tilde{c}\,v}{2\xi} - \frac{m}{2}\,v^2\right]
\ee
\be \label{eq:VOS2}
\frac{dv}{dt} = \frac{1}{t}(1-v^2)\left[\frac{k(v)}{\xi} - (2Ht - m)\,v\right]
\ee

\subsection{Cosmic superstrings}

The VOS modelling for cosmic strings can be extended to a multi-tension network of cosmic superstrings, yielding, for constant-tension strings, \cite{Avgoustidis:2007aa,Avgoustidis:2009ke,Pourtsidou:2010gu}
\bea\label{eq:superVOSL}
    \frac{\td L_i}{\td t}&=H(t) L_i (1+v_i^2)+\frac{1}{2}\Tilde{c}_i v_i\\
    &+\frac{1}{2}\bigg(\sum_{a,k}\Tilde{d}^k_{ia}\frac{v_{ia}\ell_{ia}}{L_i^2L_a^2}-\sum_{b,a\leq b}\Tilde{d}^i_{ab}\frac{v_{ab}\ell_{ab}}{L_a^2L_b^2}\bigg)L_i^3\, ,
\eea
\bea\label{eq:superVOSv}
    \frac{\td v_i}{\td t}=(1-v_i^2)\bigg[&\frac{k(v_i)}{L_i} -2H(t) v_i \\
     + &B\sum_{b,a\leq b} \Tilde{d}^i_{ab}\frac{v_{ab}}{v_i} \frac{\mu_a+\mu_b+\mu_i}{\mu_i}\frac{\ell_{ab} L_i^2}{L_a^2L_b^2}\bigg]\, ,
\eea
where now $\tilde{c}_i$ is the {\it self-intersection} (or loop chopping) efficiency parameter for strings of type $i$, and $\Tilde{d}^k_{ij}=\Tilde{d}^k_{ji}$ are the {\it cross-string intersection} efficiency parameters describing processes in which a string of type $i$ interacts with a string of type $j$, with relative velocity $v_{ij}$, to produce a type $k$ string segment of length $\ell_{ij}$.

Given that the interactions between different species of cosmic superstrings happen nearly instantaneously, the interaction probabilities, described by $\Tilde{d}^k_{ij}$, are not affected by the tensions varying with time. As such, the above equations can easily be adapted to describe cosmic superstring networks with time-varying tensions by promoting $\mu \to \mu(t)$ and adding the appropriate time-varying terms from Eqs~\eqref{eq:VOSL}-\eqref{eq:VOSv} to these VOS equations.

\subsection{Stability analysis}

As shown by \cite{Revello:2024gwa}, cosmic strings with time-varying tensions admit a second attractor solution in addition to the standard scaling solution where $\xi\sim \textrm{const}$. This \textit{ultra-relativistic} attractor arises for certain initial conditions and is described by the string network having $v\to 1$ and $\rho=\mu/L^2\sim1/a^4$, hence behaving as radiation. Note that this solution requires an additional modelling assumption; 
as $v\to 1$, length-contraction reduces the chopping efficiency, which is achieved by inserting a Lorentz term, $\Tilde{c}_i = \Tilde{c}_{0i}\sqrt{1-v_i^2}$ in the evolution Eqs.~\eqref{eq:superVOSL} and \eqref{eq:superVOSv}.

\begin{figure*}
    \centering
    \includegraphics[width=\textwidth]{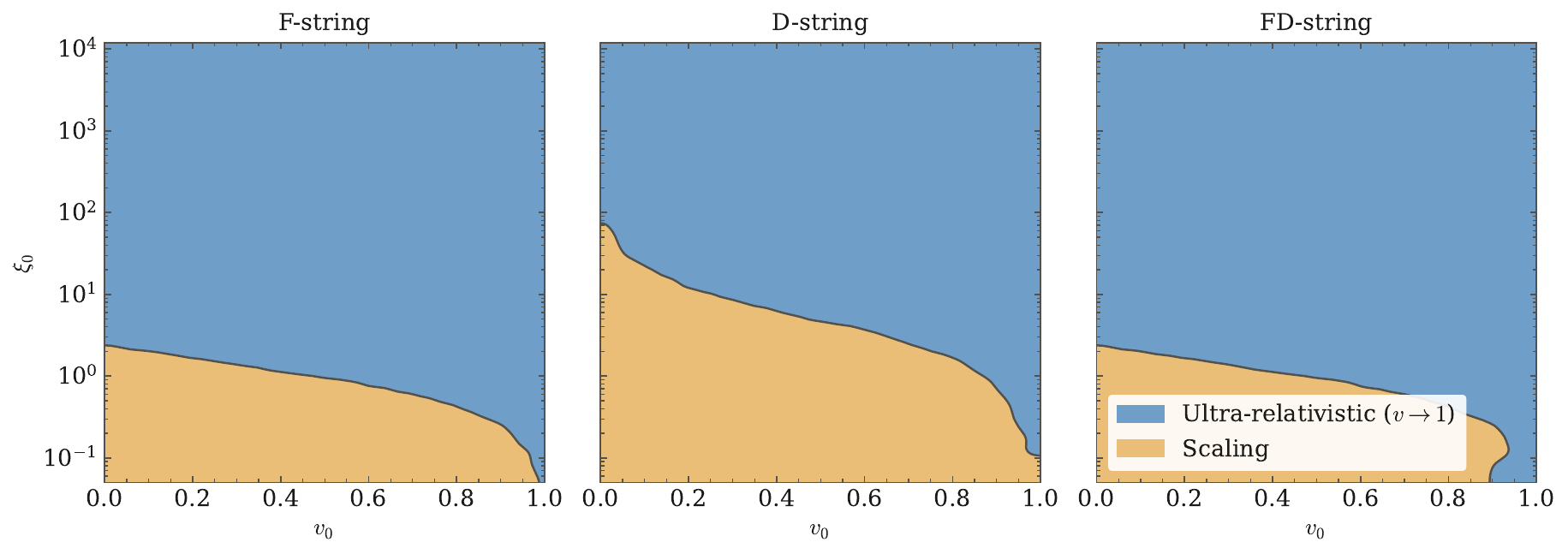}
    \caption{Example scatter plot of superstring VOS solutions for $g_s = 0.3$ and $m=1.3$ with randomly sampled initial conditions ($\xi_0,\,v_0$). The yellow region shows configurations where the superstrings reach the standard scaling solution, while blue region shows where they reach the ultra-relativistic attractor.}
    \label{fig:stability}
\end{figure*}

Due to the complicated interaction terms between string species, this analytic stability analysis is difficult to recreate for cosmic superstrings. Given that, we show numerically that the same two sets of attractor solutions can also be found in cosmic superstring networks with time-varying tensions. Figure \ref{fig:stability} shows a scatter plot of VOS solutions, for an example with $g_s = 0.3$ and $m=1.3$, with different initial conditions for $\xi_0$ and $v_0$. As can be seen from the figure, both types of attractor solutions are present. Note that this stability is very sensitive to the parameters chosen, especially the strength of the time-variance, $m$. We find that for $m<1$, almost all initial conditions lead to the standard scaling solution, while for $m>2$, almost all setups reach the ultra-relativistic attractor.

As there is no obvious causal mechanism that could yield correlation lengths much larger than $\mathcal{O}(1)$ when the network forms, and due to only considering networks with $m<1$ (to avoid percolation, discussed in more detail in Sec.~\ref{sec:GWs}), we do not expect to see the ultra-relativistic attractor solution for either cosmic strings or superstrings. In fact, the stability analysis in \cite{Revello:2024gwa} shows that, for a single string species, the ultra-relativistic attractor is not even linearly stable for $m<1$. For $m>1$, and in the case of a single string species, a semi-analytic derivation of the attraction basins for the different attractors is given in Appendix~\ref{app:atrc}.\\

\section{Gravitational waves}
\label{sec:GWs}

Throughout we adopt the one-scale loop production model, in which (super)string loops are created monochromatically at a single length set by a fixed fraction of the correlation length, $l(t_b) = \alpha\,\xi(t_b) t_b$, with the long-string network evolved using the VOS equations of Sec.~\ref{sec:modelling}. In the classification of Ref.~\cite{Auclair:2019wcv} this is Model~I. We take $\alpha = 0.37$, which with the radiation-era scaling value for cosmic strings, $\xi\simeq0.27$, corresponds to a loop size $\alpha\xi\simeq0.1\,t$, matching the characteristic loop size found in~\cite{Blanco-Pillado:2013qja}. Note that $\alpha = 0.37$ is taken to be the same for all superstring species. The loop sizes, however, vary from species to species due to differences in correlation lengths.

\subsection{Cosmic strings}

The GW energy emitted by cosmic string loops in the harmonic $j$ at observed frequency $f$ is

\be\label{eq:Omegaj}
\Omega_{\rm GW}^{(j)}(f) = \int d\ln t\;
\frac{16\pi}{3}\,\frac{\bigl[G\mu(t)\bigr]^2}{H_0^2}\,\Gamma\,\frac{j}{f}\,
n\!\left(\frac{2j\,a(t)}{f},\,t\right) a(t)^5\,t,
\ee
where $l_j(t) = 2j\,a(t)/f$ is the physical loop length, $n(l,t)$ is the number density of loops with length $l$ at time $t$ (not to be confused with the scale factor exponent $n$ in Sec.~\ref{sec:scalar-dyn}), and $\Gamma$ is the efficiency of loop decay, taken to be $50$ \cite{Blanco-Pillado:2017oxo}.

The total GW spectrum sums over harmonics with power-law weights
\be\
\Omega_{\rm GW}(f) = \sum_{j=1}^{j_{\rm max}} \frac{j^{-q}}{\zeta(q)}\,\Omega_{\rm GW}^{(1)}(f/j),
\ee
where $q=4/3$ for cusp-dominated and $q=5/3$ for kink-dominated emission \cite{Auclair:2019wcv}.

There are two modifications to Eq.~\eqref{eq:Omegaj} with respect to the constant tension case. Firstly, the tension scaling factor $(G\mu)^2$ has been promoted to be time-dependent. Secondly, the loop number density $n(l,t)$ acquires time-dependent factors. In particular, it is given by 
\be\label{eq:nloop}
    n(l,t) = \frac{\Tilde{c}v_\star}{\gamma(v_\star)\alpha\xi_\star^4 t_b^4}\frac{1}{|\td l/\td t_b|}\left(\frac{a(t_b)}{a(t)}\right)^3,
\ee
where $t_b$ is the birth time of the string loops, starred quantities denote evaluation at $t=t_b$, $\gamma$ is the Lorentz factor, and $\beta$ is the loop size fraction. As shown in \cite{Revello:2024gwa}, the length of time-dependent cosmic string loops follows (for $m\neq 2/3$)
\bea\label{eq:lloop}
    l(t) &= \left(\frac{t_b}{t}\right)^{-m/2}l_\star \\
    &-\Gamma \frac{G\mu_0}{t_0^{-m}}\frac{1}{1-3m/2}\left[t^{-m+1} - \left(\frac{t_b}{t}\right)^{-m/2}t_b^{-m+1}\right].
\eea
Combining Eq.~\eqref{eq:nloop} and Eq.~\eqref{eq:lloop} and using $l_\star = \alpha\,\xi_\star t_b$ for the loop size at birth, gives the time-dependent loop number density as
\begin{widetext}
\be\label{eq:nloop_full}
n(l,t) = \frac{\tilde{c}\,v_\star}{\gamma(v_\star)\,\alpha\,\xi_\star^4\,t_b^4}
\frac{1}{\left(\dfrac{t_b}{t}\right)^{\!-m/2}\!\Bigl[\alpha\!\left[\left(1-\dfrac{m}{2}\right)\xi_\star + t_b\,\dot{\xi}_\star\right] + \Gamma G\mu(t_b)\Bigr]}
\left(\frac{a(t_b)}{a(t)}\right)^3
\ee
\end{widetext}
Note that this treatment assumes that the cosmic string loops do not grow faster than the rate of cosmic expansion. This condition is satisfied if $2H-\frac{m}{t}>0$, which gives $m<1$ in the radiation and $m<4/3$ in the matter era. Were the loops to grow with respect to the horizon, they eventually become superhorizon and effectively become long strings, or loops could join and percolate to form long strings \cite{Conlon:2024uob} -- both effects would modify the loop number density in non-trivial ways. For this reason, we will restrict our analysis to $m<1$ for the rest of this work.

\subsection{Cosmic superstrings}

The above modelling is trivially extended to the case of cosmic superstrings by summing the GW spectra obtained from each superstring species. As discussed in detail in \cite{Avgoustidis:2025svu}, heavier strings become progressively rarer, enabling truncating the sum at the three lightest string species. Importantly, this truncation applies only to the sum over GW spectra. The VOS system itself -- the time-varying generalisation of Eqs.~\eqref{eq:superVOSL}--\eqref{eq:superVOSv} described in Sec.~\ref{sec:modelling} -- must still be solved for the full set of species, since interactions with the rarer heavy bound states feed back on the dynamics of the lighter ones.

Note that the time-dependent tension modelling for cosmic superstrings in this paper assumes that all superstring species experience the same time-variance, i.e. $m_F=m_D=\ldots\equiv m$.

\section{Time-varying tension models}
\label{sec:tension_models}

To cover more of the phenomenological space of GW spectra from (super)strings with time-varying tension, we consider, in addition to the pure power law of Eq.~\eqref{eq:tension_power}, four further scenarios: three in which the time-variation is restricted to a finite epoch, and one in which the tension depends logarithmically on time throughout. These are:
\begin{enumerate}
    \item Truncated tension evolution
    \be
        G\mu(t) =
        \begin{cases}
        G\mu_0 \left(t/t_{\rm trunc}\right)^{-m} & t < t_{\rm trunc}\\
        G\mu_0 & t \geq t_{\rm trunc}
        \end{cases}
    \ee
    \item Tension with a window of time evolution
    \be
        G\mu(t) =
        \begin{cases}
            G\mu_0 \left(t_{\rm start}/t_{\rm end}\right)^{-m} & t < t_{\rm start}\\
            G\mu_0 \left(t/t_{\rm end}\right)^{-m} & t_{\rm start} \leq t < t_{\rm end}\\
            G\mu_0 & t \geq t_{\rm end}
        \end{cases}
    \ee
    \item Tension evolution with a step-function spike
    \be\label{eq:tension_spike}
        G\mu(t) =
        \begin{cases}
            G\mu_{\rm top} & t_{\rm start} \leq t < t_{\rm end}\\
            G\mu_0 & \text{otherwise}
        \end{cases}
    \ee

\item Tension evolution with a logarithmic time dependence for all $t$
\be \label{eq:Log-time}
  G\mu(t) = G\mu_0\,
  \frac{\sqrt{1+\ln^2(t/t_\ast)}}{\sqrt{1+\ln^2(t_0/t_\ast)}},
\ee
with $t_\ast$ an initial condition preceding network formation.
\end{enumerate}
Note that, in all of these cases, $G\mu_0$ denotes the string tension at the present day.

As in the full power law model of Eq.~\eqref{eq:tension_power}, only two parts of the standard GW calculation change with
respect to constant tension -- a trivial promotion of the factor $(G\mu)^2$ to be time-dependent in Eq.~(\ref{eq:Omegaj}),
and modifications to the loop number density, Eq.~(\ref{eq:nloop}), through the loop-length solution, Eq.~\eqref{eq:lloop} in the power law case, and its Jacobian $|\td l/\td t_b|$.

For ease of writing, we abbreviate the two brackets that appear in the Jacobians of Eq.~\eqref{eq:nloop_full} as
\bea\label{eq:hprimes}
    &h_{\rm var}(t_b) \equiv \alpha\!\left[\left(1-\frac{m}{2}\right)\xi_\star
    + t_b\,\dot{\xi}_\star\right] + \Gamma\, G\mu(t_b)
    \\
    &h_{\rm c}(t_b; G\mu) \equiv \alpha\left(\xi_\star + t_b\,\dot{\xi}_\star\right)
    + \Gamma\, G\mu,
\eea
where $h_{\rm var}$ is the time-varying form of the bracket, as seen in Eq.~\eqref{eq:nloop_full}, and $h_{\rm c}$ is its standard $m=0$ form (with the appropriate constant tension).

\subsection{Truncated tension model}

While the tension varies ($t < t_{\rm trunc}$), the loop length follows
Eq.~(\ref{eq:lloop}) with $t_0 \to t_{\rm trunc}$. Once the tension freezes, the loops decay linearly. A loop born before $t_{\rm trunc}$ and observed at some later time therefore has a length
\be\label{eq:lloop_trunc}
    l(t) = l(t_{\rm trunc}) - \Gamma\,G\mu_0\,(t - t_{\rm trunc}),
    \qquad t_b < t_{\rm trunc} < t,
\ee
with $l(t_{\rm trunc})$ given by Eq.~(\ref{eq:lloop}). Differentiating the
loop-length solution with respect to $t_b$ in each regime gives for the loop density function Eq.~\eqref{eq:nloop} 
\bea\label{eq:nloop_trunc}
    n(l,t) &= \frac{\,\tilde{c}\,v_\star}
    {\gamma(v_\star)\,\alpha\,\xi_\star^4\, t_b^4}
    \left(\frac{a(t_b)}{a(t)}\right)^{3}\\
    &\times
    \begin{cases}
        \dfrac{1}{h_{\rm c}(t_b;\,G\mu_0)} & t_b \geq t_{\rm trunc},\\[2ex]
        \dfrac{(t/t_b)^{-m/2}}{h_{\rm var}(t_b)} & t \leq t_{\rm trunc},\\[2ex]
        \dfrac{(t_{\rm trunc}/t_b)^{-m/2}}{h_{\rm var}(t_b)} & t_b < t_{\rm trunc} < t.
    \end{cases}
\eea
The first two cases are identical to the constant-tension and full-model results, Eq.~\eqref{eq:nloop} and Eq.~\eqref{eq:nloop_full} respectively. In the third, the linear decay of Eq.~(\ref{eq:lloop_trunc}) is independent of $t_b$-derivatives beyond $l(t_{\rm trunc})$, so the $(t_b/t)^{-m/2}$ factor of the full model simply freezes at $t = t_{\rm trunc}$.

\subsection{Window of time evolution model}

Writing $G\mu_{\rm start} = G\mu_0\,(t_{\rm start}/t_{\rm end})^{-m}$ for the
early-time plateau tension, the loop length follows from integrating the decay
law piecewise -- linear decay wherever the tension is constant, Eq.~(\ref{eq:lloop})
inside the window, and $l$ matched continuously at $t_{\rm start}$ and
$t_{\rm end}$. A loop born at $t_b$ and sourcing GWs at time $t$ can straddle the window
in six distinct ways, and differentiating each of the resulting solutions with
respect to $t_b$ gives
\bea\label{eq:nloop_window}
    n(l,t) &= \frac{\,\tilde{c}\,v_\star}
    {\gamma(v_\star)\,\alpha\,\xi_\star^4\, t_b^4}
    \left(\frac{a(t_b)}{a(t)}\right)^{3}
    \\
    &\times
    \begin{cases}
        \dfrac{1}{h_{\rm c}(t_b;\,G\mu_0)} & t_b \geq t_{\rm end},\\[2ex]
        \dfrac{(t/t_b)^{-m/2}}{h_{\rm var}(t_b)} & t_{\rm start}\leq t_b,\ t \leq t_{\rm end},\\[2ex]
        \dfrac{(t_{\rm end}/t_b)^{-m/2}}{h_{\rm var}(t_b)} & t_{\rm start}\leq t_b < t_{\rm end} < t,\\[2ex]
        \dfrac{1}{h_{\rm c}(t_b;\,G\mu_{\rm start})} & t_b < t_{\rm start},\ t \leq t_{\rm start},\\[2ex]
        \dfrac{(t/t_{\rm start})^{-m/2}}{h_{\rm c}(t_b;\,G\mu_{\rm start})} & t_b < t_{\rm start} < t \leq t_{\rm end},\\[2ex]
        \dfrac{(t_{\rm end}/t_{\rm start})^{-m/2}}{h_{\rm c}(t_b;\,G\mu_{\rm start})} & t_b < t_{\rm start},\ t > t_{\rm end}.
    \end{cases}
\eea
In each case the denominator is $h_{\rm var}$ if the loop is born inside the
window of time-variation and $h_{\rm c}$, evaluated at the local tension, if it is
born during a plateau, while the prefactor
$\bigl(\min(t,t_{\rm end})/\max(t_b,t_{\rm start})\bigr)^{-m/2}$ spans the portion
of $[t_b,t]$ overlapping the window, reducing to unity when there is no overlap.

\subsection{Top-hat model}

Since the tension is a piecewise chain of three constant tensions, every phase of the loop evolution is linear decay,
\be\label{eq:lloop_spike}
    l(t) = l_\star - \Gamma\!\int_{t_b}^{t} G\mu(t')\,\td t',
\ee
with $G\mu(t')$ from Eq.~\eqref{eq:tension_spike}. Since the integral depends on
$t_b$ only through its lower limit, the Jacobian is simply the constant-tension
case evaluated with the tension at birth:
\be\label{eq:nloop_spike}
    n(l,t) = \frac{\,\tilde{c}\,v_\star}
    {\gamma(v_\star)\,\alpha\,\xi_\star^4\, t_b^4}\,
    \frac{1}{h_{\rm c}\big(t_b;\,G\mu(t_b)\big)}
    \left(\frac{a(t_b)}{a(t)}\right)^{3}.
\ee

\subsection{Axion kination model}
Motivated by Eq.~(\ref{eq:logaxion}), we also consider the evolution defined in Eq.~\eqref{eq:Log-time}.
Given $t_\ast$ is an initial condition preceding network formation, it implies
$\ln(t/t_\ast)\ge 0$ throughout and the tension grows asymptotically as
$\mu\propto\ln(t/t_\ast)$. The effective index follows analytically,
\be
  -m(t) = \frac{\td\ln\mu}{\td\ln t} = \frac{\ln(t/t_\ast)}{1+(\ln(t/t_\ast))^2},
\ee
which is bounded by $1/2$. This model therefore satisfies the percolation and loop-growth bounds of Sec.~\ref{sec:GWs} automatically, rather than by restriction. The loop number density retains the form of Eq.~\eqref{eq:nloop_full}, but since $\mu$ is no longer a power law the substitution $m\to m(t)$ alone is not sufficient: the growth factor $(t_b/t)^{-m/2}$ becomes $[\mu(t_b)/\mu(t)]^{1/2}$, the gravitational-wave term, analytic for a power law, becomes $\int_{t_b}^{t}[\mu(t')/\mu(t_0)]^{3/2}\,\td t'$, and $m\to m(t_b)$ in the Jacobian. We evaluate all three numerically.

In all models the birth time $t_b(l,t)$ is obtained by numerically
inverting the corresponding loop-length solution, after which the
corresponding loop number density is inserted into
Eq.~(\ref{eq:Omegaj}) with the appropriate $G\mu(t)$. All other elements of
the calculation are unchanged from the constant-tension case.

\section{CMB anisotropies}
\label{sec:CMB}

\subsection{Unequal-time correlators with varying tension}

Cosmic strings source CMB anisotropies actively, so their contribution is
fixed by the unequal-time correlators (UETCs) of the string stress-energy,
$\langle \Theta_{\mu\nu}(k,\tau_1)\Theta_{\rho\sigma}(k,\tau_2)\rangle$.
We work in the analytic UETC framework of
\cite{Avgoustidis:2012gb,Charnock:2016nzm, Pogosian:1999np,Raidal:2026cpb}, in which the correlators of a
Nambu--Goto network are written as closed-form integrals over the VOS
variables $\xi(\tau)$ and $v(\tau)$, and we evolve the resulting source terms
with the \texttt{CAMBactive} extension of \texttt{CAMB}
\cite{Raidal:2026cpb, Raidal_CAMBactive_CAMB_extension_2026}.

The string tension enters the analytic UETCs only through an overall
$\mu^2$ scaling. A time-varying tension therefore modifies the correlators by
a single multiplicative factor,\footnote{Here we assume that the wiggliness $\beta$ of the string, which defines an effective tension $\tilde{\mu}=\beta \mu$ (usually called $\alpha$~\cite{Pogosian:1999np}), is not significantly modified by the time-variation of the tension.
A preliminary analysis~\cite{Pivetti_thesis} based on the Polchinski-Rocha description of small scale structure~\cite{Polchinski:2006ee} suggests that this is the case, although conclusive evidence would require numerical simulations.}
\be\label{eq:uetc_scaling}
    \langle\Theta(k,\tau_1)\Theta(k,\tau_2)\rangle
    = \left(\frac{t(\tau_1)\,t(\tau_2)}{t_0^{2}}\right)^{-m}
      \big\langle\Theta(k,\tau_1)\Theta(k,\tau_2)\big\rangle\Big|_{\mu=\mu_0},
\ee
where $t(\tau)$ is the physical time corresponding to conformal time $\tau$
and $t_0$ its value today, consistent with the parametrisation of
Eq.~\eqref{eq:tension_power}. Note that the VOS solutions entering the correlators are now those of the time-varying equations, Eqs.~\eqref{eq:VOS1}--\eqref{eq:VOS2}.

\subsection{Eigen decomposition and source reconstruction}

Following \cite{Charnock:2016nzm,Raidal:2026cpb}, we discretize the UETCs on a grid of $k\tau$ and construct weighted matrices
$\widetilde{C} = W^{1/2} C\, W^{1/2}$, with
\be\label{eq:weight}
    W(\tau_i,\tau_j) = \big(k^2\tau_i\tau_j\big)^{\gamma}
    \sqrt{\tau_i\tau_j}\,
    \left(\frac{t_i t_j}{t_0^{2}}\right)^{m},
\ee
where $\gamma$ is a weighting exponent (following \cite{Charnock:2016nzm,Avgoustidis:2012gb}, we take $\gamma=1/4$) and the time-dependent factor removes the tension scaling of Eq.~\eqref{eq:uetc_scaling}, so that the weighted correlator (and hence its eigenvectors) remain smooth. The scalar sector is diagonalised as a single $2N\times 2N$ system in the coupled $(\Theta_{00},\Theta_{S})$ basis, while the vector and tensor sectors are diagonalised independently.

Each eigenmode $p$, with eigenvalue $\lambda_p$ and eigenvector $u_p$, defines
an effective source,
\be\label{eq:source}
    S_p(k,\tau) = \sqrt{|\lambda_p|}\;\frac{u_p(k\tau)}{w(k\tau)},
    \qquad w \equiv \sqrt{\mathrm{diag}\,W},
\ee
which is passed to the Einstein-Boltzmann solver of \texttt{CAMBactive} \cite{Raidal_CAMBactive_CAMB_extension_2026} as a source term. Since distinct
eigenmodes are uncorrelated, the total angular power spectrum is the incoherent sum
\be\label{eq:clsum}
    C_\ell = \sum_{p} C_\ell^{(p)} .
\ee
For cosmic superstrings the procedure is identical, carried out separately for
each string species and the total is obtained by summing the species spectra
weighted by the appropriate tensions, as in Sec.~\ref{sec:GWs}. Analogously to the GW case, only the anisotropies of the three lightest string species need to be explicitly computed and included in the total.

\subsection{Numerical setup}

We tabulate the correlators on $100$ logarithmically spaced wavenumbers in
$k\in[10^{-6},10]\,\mathrm{Mpc}^{-1}$ and $1024$ points in
$k\tau\in[10^{-4},9\times10^{3}]$, retaining the $128$ largest eigenmodes of
each sector. These resolution parameters are determined based on the following criterion. For a given resolution we form the ratio
\be
    f = \frac{\mathrm{SNR}\!\left[\Delta C_\ell\right]}
             {\mathrm{SNR}\!\left[C_\ell\right]},
    \qquad
    \mathrm{SNR}^2[X_\ell] = \sum_\ell \left(\frac{X_\ell}{N_\ell}\right)^2,
\ee
where $\mathrm{SNR}[\,\cdot\,]$ is a functional of an arbitrary angular power spectrum $X_\ell$, evaluated here at $X_\ell=C_\ell$ and $X_\ell=\Delta C_\ell$. $\Delta C_\ell$ is the difference from a high-resolution reference spectrum ($1200\times1200\times 500$) and $N_\ell$ is the total noise, combining cosmic variance with the ACT DR6 noise \cite{AtacamaCosmologyTelescope:2025blo}. Both the numerator and the denominator scale as $(G\mu_0)^2$, so $f$ is independent of the tension and can be evaluated separately at each $m$. At the tension where the model is
excluded at $2\sigma$ one has $\mathrm{SNR}[C_\ell]=2$ by definition, so
requiring the resolution error to be undetectable, $\mathrm{SNR}[\Delta
C_\ell]<1$, is equivalent to $f<1/2$. This condition is satisfied, on the interval $m\in[-1,1]$, for a spectrum derived from UETCs computed on a grid of $1024\times1024\times100$, which enables keeping the CMB calculations computationally feasible.

Sampling the posterior directly would require a full UETC computation per
likelihood evaluation, making a direct Monte Carlo simulation unfeasible. Following \cite{Raidal:2026cpb}, we therefore train a neural network emulator mapping the model parameters $\{\beta, \tilde{c}, m, H_0, \Omega_m, \Omega_r, \Omega_k\}$ (with $g_s$ and $w$ added for superstrings) onto the flattened $TT$, $EE$, $BB$ and $TE$ spectra using a training set of $1000$ spectra for both cosmic strings and superstrings.

\begin{figure*}
    \centering
    \includegraphics[width=\textwidth]{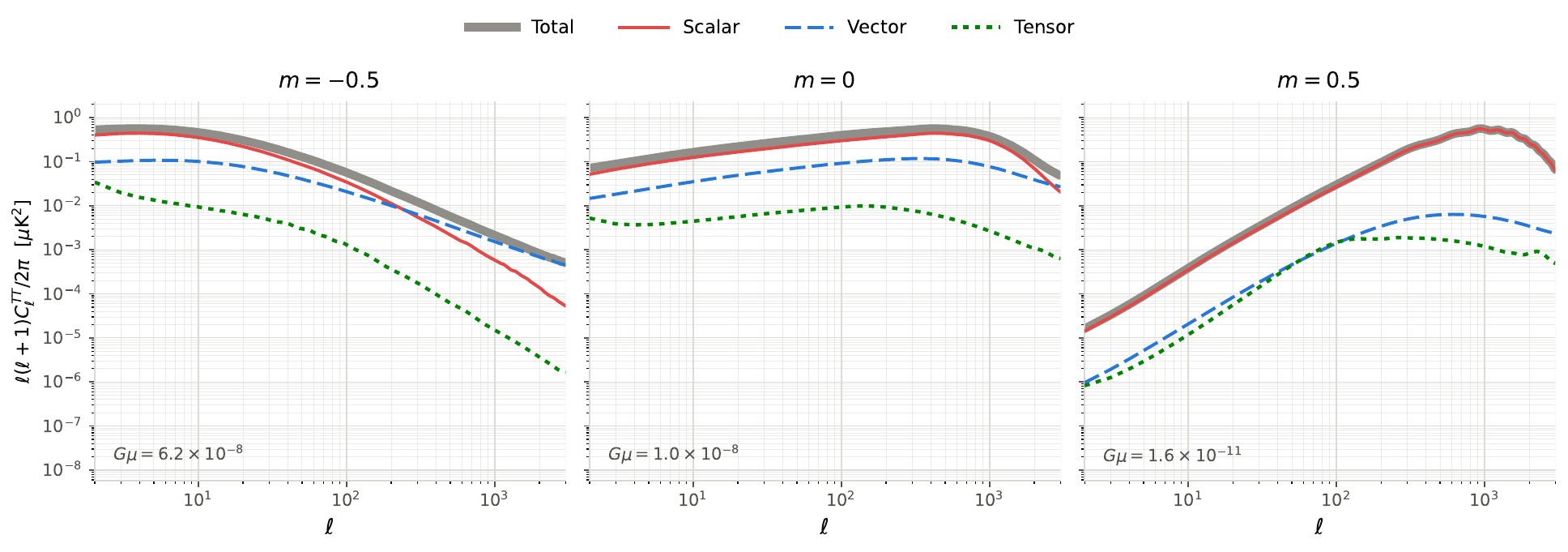}
    \caption{$TT$ angular power spectrum sourced by cosmic strings with
    time-varying tension given by Eq.~\eqref{eq:tension_power}, decomposed into scalar, vector and tensor
    contributions, for three values of the tension index $m$. Each panel is
    normalised so that the peak of the total amplitude matches the $m=0$ case at
    $G\mu_0=10^{-8}$; the tension required to do so is annotated in each panel.}
    \label{fig:CMBspectra}
\end{figure*}

Figure~\ref{fig:CMBspectra} shows the $TT$ spectrum and its scalar, vector and
tensor contributions for representative values of $m$ at fixed present-day
tension. The time-dependent modification with respect to the constant-tension case consists of two parts -- a scaling of the amplitude and a tilt to the spectrum. For $m>0$, where the tension was
larger in the past, small-scale modes (which are sourced early) are enhanced
relative to the constant-tension case and the spectrum tilts upwards, while the
overall amplitude at fixed $G\mu_0$ grows by several orders of magnitude.
For $m<0$ the opposite holds, with the high-$\ell$ signal suppressed. These results are consistent with~\cite{Yamaguchi:2005gp,Ichikawa:2006rw,Martins:2018dqg}, although different VOS equations and computational methods were used in this work.

\section{Results}
\label{sec:results}

\subsection{CMB constraints}

We fit the all-time power-law time-variance model, Eq.~\eqref{eq:tension_power}, to \textit{Planck} 2018 low-$\ell$ \cite{Planck:2018vyg} and ACT DR6 \cite{AtacamaCosmologyTelescope:2025blo} high-$\ell$ polarization data, with the MCMC sampling done using \texttt{Cobaya} \cite{Torrado_2021}. In addition to the $\Lambda$CDM parameters we vary the (super)string tension with a log-uniform prior on $\log_{10}G\mu(t_0)$($\log_{10}G\mu_F(t_0)$)$\in[-15, -4]([-16, -5])$, the tension index with a flat prior $m\in[-1,1]$, and the UETC and VOS shape parameters $\beta=1.7\pm 0.3$ and $\tilde{c}=0.23\pm 0.04$ with Gaussian priors centred on their simulation-derived values. Here we make the assumption that these values do not significantly deviate from the simulation-derived priors due to the time-varying tension. For cosmic superstrings, the string coupling and volume parameter are also varied, with flat priors -- $g_s\in[0.01,0.9]$ and $w\in[0.1,1]$. It is important to note that, as discussed in detail in \cite{Raidal:2026cpb}, the choice of priors significantly impacts the tension constraints obtained from CMB analyses, and only like-for-like comparisons are to be made between different analysis results.

\begin{figure*}
    \centering
    \includegraphics[width=\textwidth]{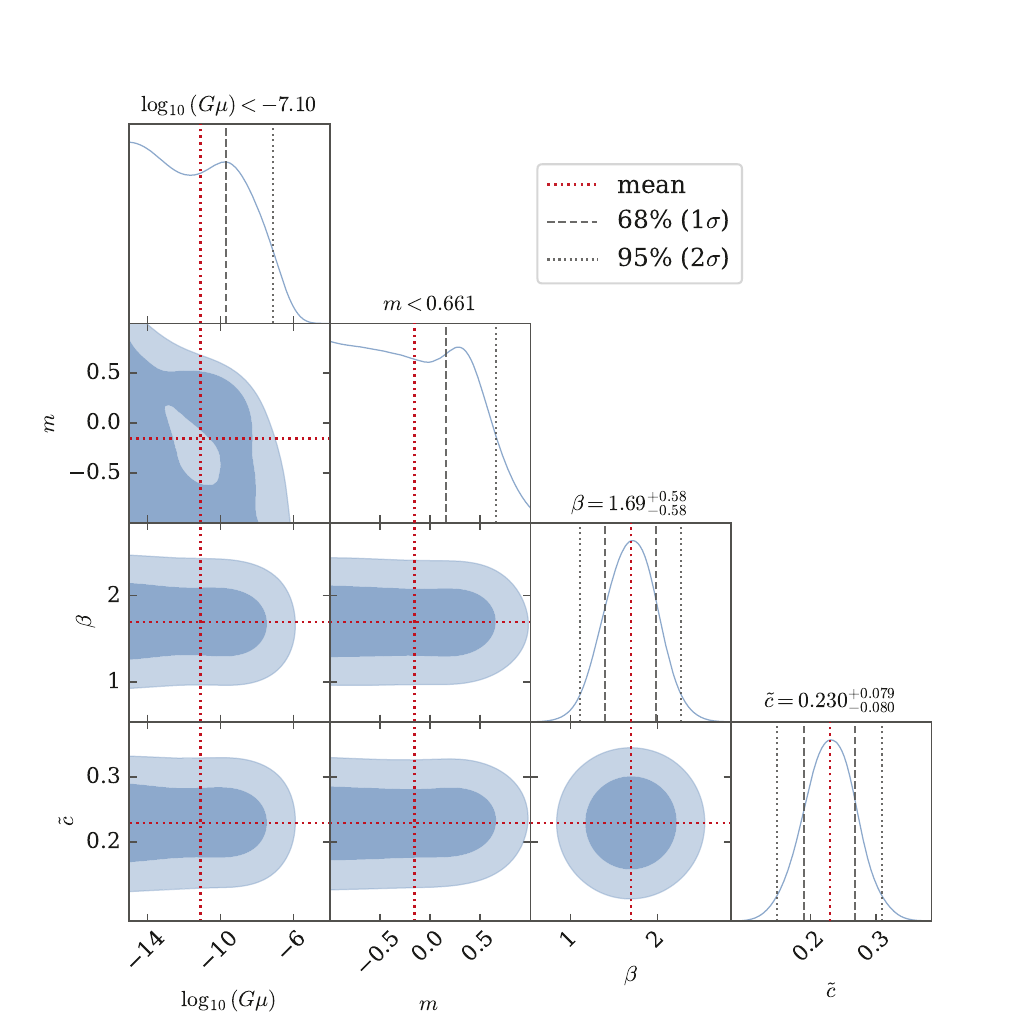}
    \caption{Corner plot showing the posterior marginals for the MCMC analysis of cosmic strings to Planck+ACT CMB polarization data. The values quoted above each 1D marginal are the 95\% confidence levels or bounds. Note that Gaussian priors were used for $\beta$ and $\tilde{c}$ and the posteriors are prior-dominated, not determined by the data.}
    \label{fig:CMBcornerCS}
\end{figure*}

\begin{figure*}
    \centering
    \includegraphics[width=0.9\textwidth]{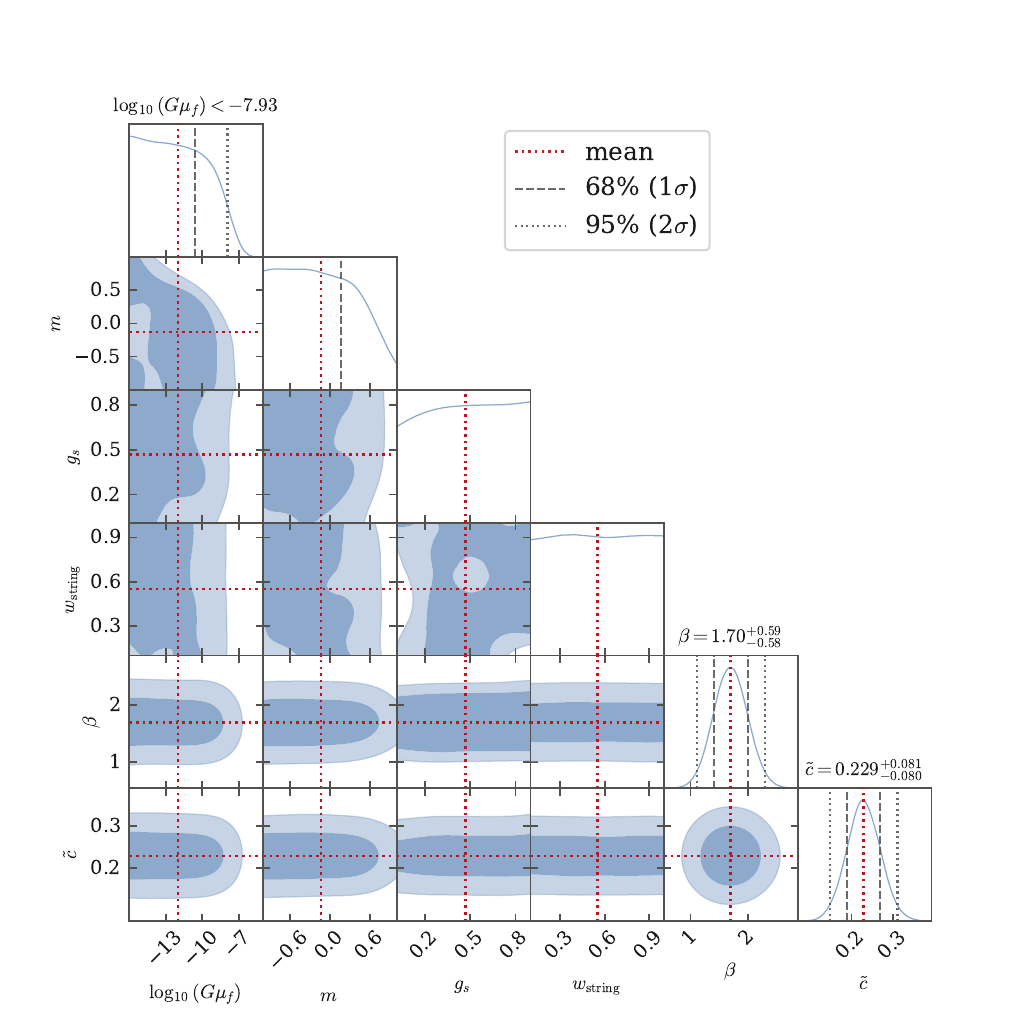}
    \caption{Corner plot showing the posterior marginals for the MCMC analysis of cosmic superstrings to Planck+ACT CMB polarization data. The values quoted above each 1D marginal are the 95\% confidence levels or bounds. Where a bound cannot be placed, no value is quoted. Note that Gaussian priors were used for $\beta$ and $\tilde{c}$ and the posteriors are prior-dominated, not determined by the data.}
    \label{fig:CMBcornerSS}
\end{figure*}

The resulting posteriors are shown in Figs.~\ref{fig:CMBcornerCS} for cosmic strings and \ref{fig:CMBcornerSS} for cosmic superstrings. As in the constant-tension case, cosmic strings do not improve the fit to CMB data, so the analysis yields an upper bound on the tension rather than a detection, and the remaining model parameters are only weakly constrained. We find $\log_{10}G\mu(t_0) < -7.10$ at $95\%$ confidence, to be compared with $\log_{10}G\mu(t_0) < -7.43$ for a constant-tension network \cite{Raidal:2026cpb}. A loose bound on the tension can be placed at $m < 0.66$ at $95\%$ -- networks whose tension was substantially larger in the past overproduce small-scale anisotropy and are excluded, while $m<0$ is essentially unconstrained because it suppresses the signal. Note that the tension constraint is slightly weakened with respect to the constant case. This is a result of the Bayesian volume of $m<0$, whose small tension at early times allows for a higher present-day tension.

The posteriors for cosmic superstrings follow the same qualitative picture as those for cosmic strings, although the added unconstrained flat-priored parameters $g_s$ and $w$ further complicate constraining the model. We find $\log_{10}G\mu_F(t_0) < -7.93$ at $95\%$ confidence, compared with the constant-tension result $\log_{10}G\mu_F(t_0) < -8.14$ \cite{Raidal:2026cpb}. For superstrings, however, $m$ remains unconstrained at $95\%$ confidence due to the freedom allowed by the additional unconstrained parameters.

We conclude that CMB data constrains time-varying tension only weakly, and
mainly through a small effect on the inferred tension bound. We expect this to hold for any tension histories which do not imprint oscillatory features on the spectra and therefore we will not consider any models of tension evolution beyond the all-time power law. 

\subsection{Gravitational Wave Spectra}
\label{sec:GWresults}

We fit each of the tension histories of Sec.~\ref{sec:tension_models}
to the NANOGrav 15-year dataset \cite{NANOGrav:2023gor} and the LIGO--Virgo--KAGRA O3 isotropic search non-detection threshold \cite{KAGRA:2021kbb}. For the PTA data we use the \texttt{ceffyl} representation
\cite{Lamb:2023jls} of the NANOGrav free spectrum, in which the posterior of
the timing-residual amplitude is provided as a kernel density estimate (KDE) in each
frequency bin. A model spectrum is converted to the per-component residual
\be\label{eq:sigma0}
    \sigma_0(f) = \frac{H_0}{f^{5/2}}
    \sqrt{\frac{\Omega_{\rm GW}(f)}{8\pi^4 T_{\rm obs}}},
\ee
with $T_{\rm obs}$ the observation time, and the log-likelihood is the sum over
the $N_f=14$ lowest bins of the corresponding KDE evaluations, 
\be\label{eq:lnLNG}
    \ln\mathcal{L}_{\rm NG15}
    = \sum_{i=1}^{N_f} \ln p_i\!\left[\log_{10}\sigma_0(f_i)\right].
\ee

The LVK data enter as a non-detection. We use the O3 broadband sensitivity
$\sigma_{\rm GW}(f)$ of \cite{KAGRA:2021kbb}, defined as the square root of the
variance of the cross-correlation statistic and therefore giving the
sensitivity of the search to narrowband features at each frequency. Taking it
as a per-frequency uncertainty on a vanishing signal, the log-likelihood is \cite{Mandic:2012pj} 
\be\label{eq:lnLLVK}
    \ln\mathcal{L}_{\rm LVK}
    = -\frac{1}{2}\sum_{f}
      \left(\frac{\Omega_{\rm GW}(f)}{\sigma_{\rm GW}(f)}\right)^{2},
\ee
where the sum runs over the frequency range covered by the search. Using
$\sigma_{\rm GW}(f)$ rather than the power-law integrated curve avoids
double-counting the frequency integration, which the sum above performs
explicitly. The reference value is zero rather than a measurement, so this term
penalises any predicted power and is maximised by a vanishing signal. The total likelihood is
$\ln\mathcal{L}=\ln\mathcal{L}_{\rm NG15}+\ln\mathcal{L}_{\rm LVK}$.

\begin{figure*}
    \centering
    \includegraphics[width=0.8\textwidth]{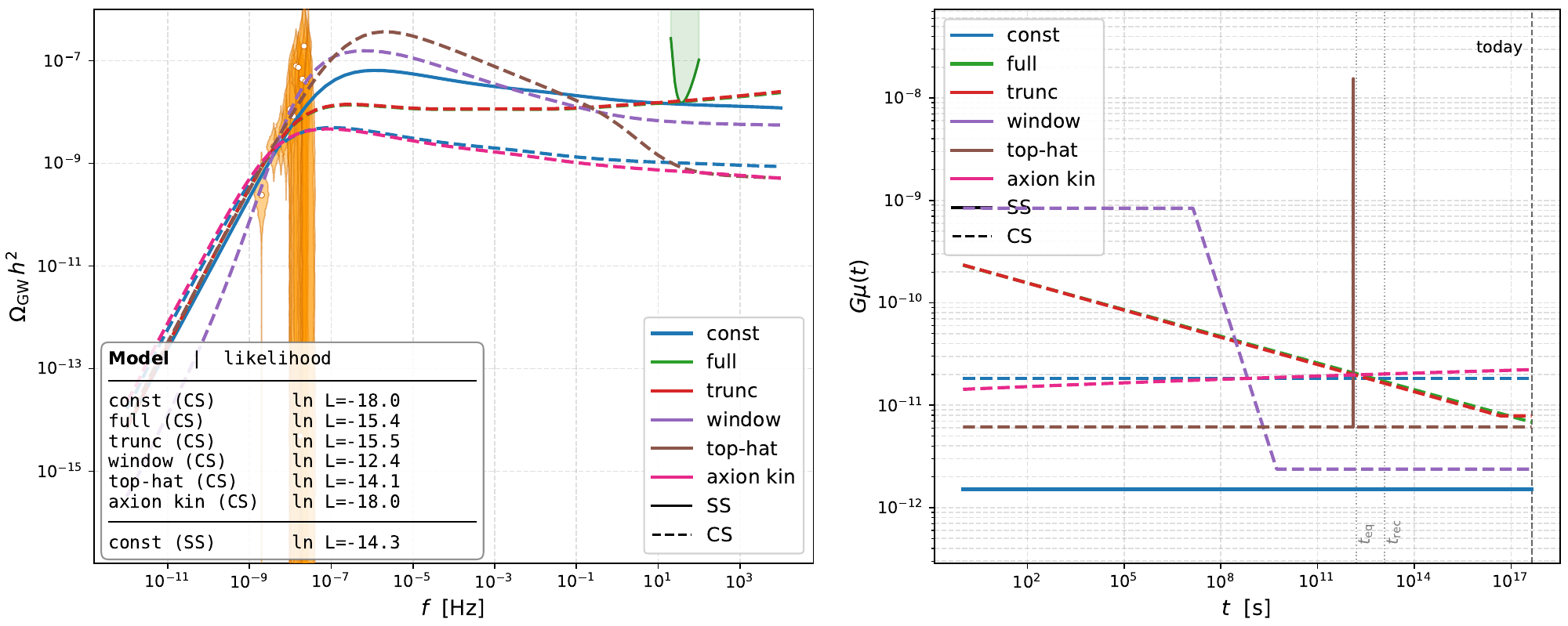}
    \caption{The best-fit spectra of each of the time-varying models for cosmic strings (dashed lines) with the constant tension cosmic superstring fit shown for comparison as a solid line. The NANOGrav 15-year data bins are shown as orange violins, while the LVK O3 non-detection bound is shown as a green shaded region.}
    \label{fig:bestfitsGW}
\end{figure*}

\begin{figure*}
    \centering
    \includegraphics[width=0.8\textwidth]{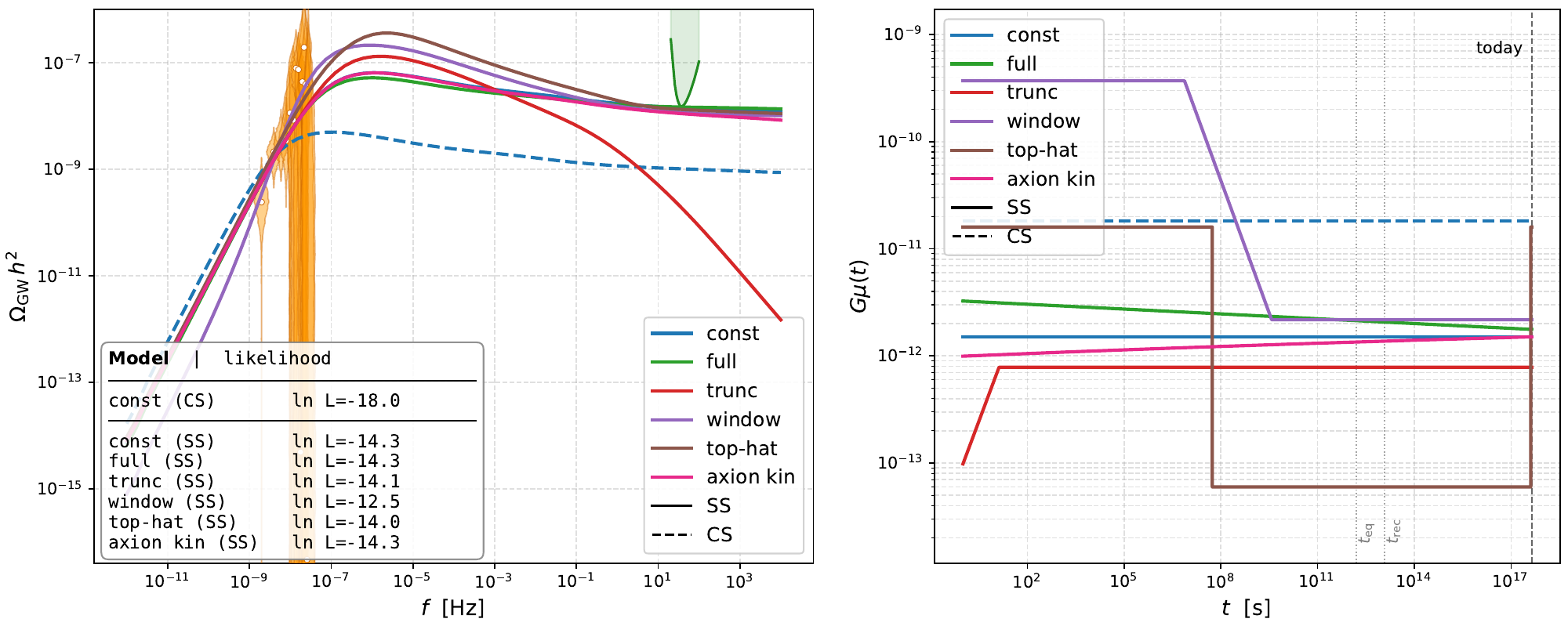}
    \caption{The best-fit spectra of each of the time-varying models for cosmic superstrings (solid lines) with the constant tension cosmic string fit shown for comparison as a dashed line. The NANOGrav 15-year data bins are shown as orange violins, while the LVK O3 non-detection bound is shown as a green shaded region.}
    \label{fig:bestfitsGWSS}
\end{figure*}

Posteriors are sampled with \texttt{emcee} \cite{Foreman-Mackey:2012any}.
We adopt flat priors throughout: $\log_{10}G\mu(t_0)\in[-14,-6]$ (and
likewise for $\log_{10}G\mu_{\rm top}$ in the top-hat model), $m\in[-3,1]$,
and $\log_{10}(t_{\rm start}/{\rm s}),\ \log_{10}(t_{\rm end}/{\rm s})\in[0,18]$
subject to $t_{\rm start}<t_{\rm end}$. For superstrings we additionally vary
$\log_{10}g_s$ and $\log_{10}w$, both flat on $[-2,0]$. The upper limit on $m$
reflects the loop-growth condition of Sec.~\ref{sec:GWs}. We assume that all loops are born with a size which is a constant fraction of the network correlation length, i.e. $\alpha=0.37$ \cite{Blanco-Pillado:2013qja}, and that the loop chopping efficiency takes its simulation-derived central value $\tilde{c}=0.23$\cite{Martins:2000cs}. For the windowed and top-hat models we sample the interval in terms of its
geometric centre $t_{\rm c}$ and its width in decades $\Delta$, rather than by its endpoints,
\be
    \log_{10}t_{\rm start} = \log_{10}t_{\rm c} - \tfrac{1}{2}\Delta,
    \qquad
    \log_{10}t_{\rm end} = \log_{10}t_{\rm c} + \tfrac{1}{2}\Delta,
\ee
or equivalently $$t_{\rm c}=\sqrt{t_{\rm start}t_{\rm end}}, \qquad~ 
\Delta=\log_{10}(t_{\rm end}/t_{\rm start}).$$
Figures~\ref{fig:bestfitsGW} and \ref{fig:bestfitsGWSS} show the maximum-likelihood spectra for each model, the likelihood values and the corresponding tension histories. For cosmic superstrings the constant-tension network is already a good fit to the PTA data, and the time-varying models do little to improve on it. As the constant tension best-fit already sits close to the LVK bound, the full-time power-law and truncated models both prefer tension histories very close to the constant tension case. Similarly, the axion kination model, where tension varies slowly as a logarithm of time, is nearly indistinguishable from the constant tension case. The top-hat and window models, in contrast, yield spectra that differ significantly from the constant-tension baseline in the frequencies between the PTA and LVK data, while the spectral shape within the NG15 bins remains similar to the constant-tension fit.

For cosmic strings, a constant-tension spectrum is too flat to reproduce the PTA signal well, and the additional freedom of a time-varying tension yields better fits. All time-varying fits, except the axion kination model, improve significantly on the constant-tension likelihood, with most models being comparable in likelihood and in PTA-range spectral shape to the constant-tension superstring fit, while the window model yields an even better fit. This makes time-varying cosmic strings a competitive candidate for the current stochastic GW background data, one of the main results of this work.

To assess whether the improvement in fit justifies the additional parameters introduced by a time-varying string tension, we compute for each model the Bayesian information criterion,
\be\label{eq:BIC}
    \mathrm{BIC} = k\ln N - 2\ln\mathcal{L}_{\rm max},
\ee
where $k$ is the number of sampled parameters and $N$ the number of data
points. We take $N=15$ -- the $14$ NANOGrav frequency bins, plus one for the LVK constraint. Table~\ref{tab:BIC} collects the results, quoted relative to the constant-tension network of the same type. We report both $\Delta\chi^2=-2\Delta\ln\mathcal{L}_{\rm max}$, the raw improvement in fit,
and $\Delta\mathrm{BIC}$, which is that improvement after the $k\ln N$ penalty.

\begin{table}[t]
\centering
\begin{tabular}{llccccc}
\hline\hline
Model & Type & $k$ & $\ln\mathcal{L}_{\max}$ & $\Delta\chi^2$ & BIC & $\Delta$BIC \\
\hline
const     & CS & 1 & $-18.00$ & ---       & $38.70$ & ---     \\
axion kination & CS & 2 & $-17.99$ & $-0.02$   & $41.39$ & $+2.68$ \\
window    & CS & 4 & $-12.36$ & $-11.28$  & $35.54$ & $-3.16$ \\
full      & CS & 2 & $-15.36$ & $-5.28$   & $36.13$ & $-2.57$ \\
top-hat   & CS & 4 & $-14.12$ & $-7.74$   & $39.08$ & $+0.38$ \\
trunc     & CS & 3 & $-15.53$ & $-4.94$   & $39.18$ & $+0.47$ \\
\hline
const     & SS & 3 & $-14.28$ & ---       & $36.68$ & ---     \\
axion kination & SS & 4 & $-14.28$ & $+0.02$   & $39.40$ & $+2.72$ \\
full      & SS & 4 & $-14.34$ & $+0.14$   & $39.52$ & $+2.84$ \\
window    & SS & 6 & $-12.53$ & $-3.49$   & $41.31$ & $+4.63$ \\
trunc     & SS & 5 & $-14.12$ & $-0.30$   & $41.79$ & $+5.11$ \\
top-hat   & SS & 6 & $-14.01$ & $-0.53$   & $44.27$ & $+7.59$ \\
\hline\hline
\end{tabular}

\caption{Model comparison using Bayesian information criteria. Each $\Delta$BIC is quoted relative to the constant-tension baseline of the same string type. Negative values favour the model over that baseline.}
\label{tab:BIC}
\end{table}

As the superstring models saw only minor changes in likelihood with respect to the constant tension case, whilst adding 1-3 extra parameters, the models are unsurprisingly disfavoured by the BIC. For cosmic strings, by contrast, every
time-varying power law model is either nearly neutral ($\Delta BIC\approx0$) or favoured over the constant tension baseline, with the window model being most preferred at $\Delta\mathrm{BIC}=-3.16$. The one exception is the axion kination model, whose spectra and likelihood are nearly indistinguishable from the constant tension case, while adding an extra parameter, leading to a disfavoured fit to data. We note that, although the best-fit time-varying cosmic string model (window) improves on the likelihood with respect to the constant-tension cosmic superstring model, it also adds an extra parameter and the two models are approximately equally preferred by the data.

Finally, we look at the spectral shapes themselves. As all of the tension models considered, except the top-hat, evolve smoothly, there are no uniquely distinctive signatures present in the GW signal. The full, truncated, and axion kination models largely only provide minor tilts to the radiation-era plateau in the $\rm mHz$-to-$\rm kHz$ range, with the cosmic string full model best-fit being notable for giving a slightly upturned plateau, which is not generic to standard string spectra and which would be a feature to look for in future LISA data \cite{LISACosmologyWorkingGroup:2022jok}. The windowed and top-hat models produce the most distinctive spectra, both generating a broad-frequency change in the spectrum at frequencies intermediate between the PTA and LVK bands, in the range accessible to LISA. None of these features, however, are particularly distinct -- changes to the radiation-era plateau can be described through modified cosmological evolution \cite{Ghoshal:2025tlk, Ellis:2023tsl}, while the GW spectral shape depends on loop sizes and other modelling assumptions, and could likely be explained by stringy models not involving time-varying tensions.

It is also important to not place too much weight on the best-fit spectra and rather look at the full posterior marginals, presented as corner plots in Appendix \ref{app:corners}, where the priors for each marginal are shown as dashed lines. We see that, generally, the tensions are well-constrained by the data, as is the tension-scaling $m$ in the full time-varying models, while the timing parameters (start and end of the time-variance) as well as the superstring coupling and volume parameter, are largely prior-dominated and not well-constrained. In the truncated and windowed models, $m$ itself is only weakly constrained, retaining most of its prior width. The potentially interesting structures seen in the best-fit are therefore not all equally robust. For the cosmic string full time varying model, $m=0.082^{+0.027}_{-0.047}$, yielding $P(m>0)=0.93$, i.e. an upturned plateau is preferred in $\sim93\%$ of the possible realisations, though the tilt itself is mild. For the truncated model, by contrast, the upturn seen in the best-fit reflects an essentially unconstrained $m$, whose marginal in fact slightly favours a downturn ($P(m>0)=0.21$).

As such, we conclude that physically reasonable time-varying tension models yield some potentially interesting spectral signatures at frequencies between current PTA and interferometer data, but these features are not unique to time-varying tensions, and those that depend on the timing of the variation are not localised or robustly required by the present data.

\section{Conclusions}
\label{sec:conclusions}

The vast majority of studies of cosmic strings and cosmic superstrings have assumed their tension, or mass per unit length, remains fixed throughout their evolution. However, recent work has highlighted the possibility that the formation of the strings, and in particular their tensions, are governed by evolving moduli fields (see for example \cite{Conlon:2024uob,Revello:2024gwa}). Time dependence could arise either because the strings themselves are composed of evolving moduli fields, or because they are directly coupled to evolving scalar fields leading to an effective time dependence in the tension. In such a scenario, it opens up the question, what would the cosmological implications and signatures be of a network of strings with time dependent tensions? The fact that strings are constantly forming loops of varying oscillating frequencies as they evolve, means that they are contributing to the Gravitational Wave signal across a wide range of frequencies. This implies that they are not only constrained by one particular GW experiment, associated with a given frequency, but by a large number of them, including, for example, Pulsar Timing Arrays such as NANOGrav, through to interferometers such as LISA and Adv LIGO. For strings with time dependent tensions, it offers the possibility of seeing evidence of the evolution through particular contributions at different frequencies, which could in principle lead to specific stringy signals. It is with this in mind, we have analysed a series of phenomenologically motivated models of evolving string tensions, obtaining their contributions to the GW spectrum and their impact on the anisotropies in the CMB, and comparing their results to the more traditional case of a fixed time independent tension. In particular, we have considered five different time varying models, motivated by different possible forms of evolution of the string tension. They range from a model in which the tension decreases or increases continuously over its evolution, through to piecewise models where the time dependence is for a finite period of time, before reaching a fixed constant value. By modifying the VOS model to allow for the time evolution in the tension, including the impact on intercommuting, we have performed an MCMC analysis of the models using both GW and CMB data. We have found that there are indeed interesting spectral signatures in the GW emission at frequencies between the current PTA and interferometer data. Unfortunately, though, we find that the types of signatures are not unique to these strings, implying they will be hard to pin down using current data. However, interestingly, we do find cosmic strings with evolving tension fit the GW data as well as other favoured models. 

The fact that string tensions could be time dependent, opens up a new way of looking for strings in the cosmos, and in this paper we have made the first attempt to quantify the impact they could have in the future. As observations improve, and we begin to get data over multiple frequencies, it makes this an exciting time to be thinking about strings in cosmology.

\begin{acknowledgements}
The work of AA, EJC and AM was supported by STFC Consolidated Grants [Grant Nos. ST/X000672/1 and ST/C004235/1]. JR was supported by an STFC studentship [Grant No.\ ST/Y509437/1]. FR acknowledges support from a junior postdoctoral fellowship of the Fonds Wetenschappelijk Onderzoek (FWO), project number 12A1Q25N.
GV is supported by a Research Fellowship from Gonville \& Caius College, Cambridge, and thanks Giacomo Pivetti for related discussions during his Master's thesis.
This work has been partially supported by STFC consolidated grant UKRI4224.
For the purpose of open access, the authors have applied a CC BY public copyright license to any Author Accepted Manuscript version arising.

\end{acknowledgements}

\section*{Appendices}
\appendix
\section{Attraction basins for the VOS model with varying tension}\label{app:atrc}
In this appendix, we provide semi-analytical formulas for the attraction basins of the two different fixed points for the VOS model with varying tension. For simplicity, let us consider the case of a single string species, as in Eqs \eqref{eq:VOS1}-\eqref{eq:VOS2}, and assume $ mn >4, n>2$, for reasons that will become clear soon. Defining
\begin{equation}
 p \equiv \frac{n-2}{mn-4} >0, \qquad \chi \equiv  \frac{\xi\,v^p}
{(1-v^2)^{\frac{p}{2}+\frac14}},
\end{equation}
the system can be recast as
\begin{equation}\label{eq:sys}
\begin{cases}
\chi'(z)=  \frac{v^p}{(1-v^2)^{\frac{p}{2}+\frac14}}
\left[
\frac{v\,\widetilde c(v)}{2}
+k(v)\left(\frac{p}{v}+\frac{v}{2}\right)
\right]  ,\\
v'(z)= (1-v^2) \left[ \frac{n-2}{n p}v
+ \frac{k(v)\,}{\chi} \frac{v^p}{(1-v^2)^{\frac{p}{2}+\frac14}} \right].
\end{cases}
\end{equation}
With this rewriting, one immediately notices that the RHS of the first equation depends only on $v$, and the fixed points are determined by its zeros. As shown in \cite{Revello:2024gwa}, there can be up to two fixed points for $0< v_s <v_u< 1$. The first one, $(v_s, \chi_s)$ is the universal, scaling fixed point, which exists for any value of $m,n$ and it is always stable. The second one,  $(v_u, \chi_u)$ is always a saddle point, and it exists for $ m n>4$. Moreover, there is also a special, relativistic fixed point, stable only if 
\begin{equation}\label{eq:rfp}
m n>\operatorname{Max}\left\{4,4+n\left(\frac{m}{2}-1\right)\right\}.
\end{equation}
Therefore, if $mn <4$ the scaling solution is the global attractor of the system. On the other hand, when Eq.~\eqref{eq:rfp} is satisfied, two attractors will coexist. Their respective attraction basins will be delimited by a separatrix $W(v)$ in the $(\chi,v)$ plane, which we assume to be smooth. Thanks to the simple form of Eq.~\eqref{eq:sys}, it obeys 
\begin{equation}
   \frac{1}{W} \frac{{\rm d} W}{{\rm d} v} =  \frac{\frac{v\,\widetilde c(v)}{2}
+k(v)\left(\frac{p}{v}+\frac{v}{2}\right)}{(1-v^2) \left[  \frac{n-2}{n p}v^{1-p} (1-v^2)^{\frac{p}{2}+\frac14}  W + k(v) \right]}.
\end{equation}
Moreover, the unknown integration constant can be determined from the fact that the $W(v)$ must contain the saddle fixed point, namely $W(v_u)=\chi_u$. In the non-relativistic limit $v \ll 1$, one can approximate the above as
\begin{equation}
    \frac{1}{W} \frac{{\rm d} W}{{\rm d} v} =  \frac{k_0 p}{v} \frac{1}{\frac{n-2}{n p}v^{1-p} W+ k_0},
\end{equation}
where $k_0\equiv k(0)= \frac{2\sqrt{2}}{\pi}$. For $p \neq 1$,
it is solved exactly by the implicit expression 
\begin{equation}
k_0 v^{p-1}
=
\frac{(n-2)(p-1)}{np}
\left[
W-C^{1/p}W^{\frac{p-1}{p}}
\right],
\end{equation}
 while for $p=1$,
\begin{equation}
    v=\frac{W}{C_2}\exp\left(\frac{n-2}{n}\frac{ W}{k_0}\right).
\end{equation}
For small $v\ll 1$, the leading order expressions are
\begin{equation}
  W(v)\simeq
\begin{cases}
C_1+\dfrac{n p^2 k_0}{(n-2)(p-1)}\,v^{p-1},
& p>1,\\
W(v)\simeq C_2 v, & p=1 \\
\left[\dfrac{(n-2)(1-p)}{k_0 np}\right]^{\frac{p}{1-p}}
C_1^{\frac{1}{1-p}}\,v^p,
& 0<p<1.
\end{cases}
\end{equation}
The values of the constant $C_1$ (or $C_2$ if $p=1$) for representative values of $m,n$ have been obtained numerically, and are reported in Table \ref{tab:val}.
\begin{table}[h]
\centering
\caption{Numerical values of $C_1$ across the $(m,n)$ grid.
Light blue denotes $p>1$, light yellow denotes $p=1$, and
light coral denotes $0<p<1$. The yellow entry is $C_2$ rather
than $C_1$. Dashes indicate $mn\leq4$, outside the applicable regime.}
\label{tab:val}
\begin{tabular}{c|cccccc}
$n\backslash m$
& $0$ & $\frac{1}{2}$ & $1$ & $\frac{3}{2}$ & $2$ & $4$ \\ \hline
$3$
& --
& --
& --
& \cellcolor{pAbove}$516$
& \cellcolor{pBelow}$4.33$
& \cellcolor{pBelow}$0.425$ \\

$4$
& --
& --
& --
& \cellcolor{pEqual}$6.92\times10^{3}$
& \cellcolor{pBelow}$2.89$
& \cellcolor{pBelow}$0.421$ \\

$6$
& --
& --
& \cellcolor{pAbove}$258$
& \cellcolor{pBelow}$8.53$
& \cellcolor{pBelow}$2.17$
& \cellcolor{pBelow}$0.412$
\end{tabular}
\end{table}
Notice how the value of $p=1$ represents a threshold between very different behaviours. For $p>1$, the low-$v$ separatrix (in the original coordinates) is approximately of the form $\xi \sim c_1/v^p$, with $c_1 \gg 1$. Therefore, the ultra-relativistic fixed point would require correlation lengths much larger than the horizon as an initial condition. On the other hand, for $0<p<1$, the separatrix can be approximated as $\xi \sim c_2 \sim \mathcal{O}(1)$. In that case, it may be possible for the network to reach the ultra-relativistic fixed point with physically sensible initial conditions. In the special case $p=1$, the separatrix is also given by $\xi \sim c_2 \sim \mathcal{O}(1)$. However, in the one example presented in the table above, $c_2 \gg 1$, which again excludes the relativistic attractor.

\section{GW corner plots}
\label{app:corners}

Here we present the full set of MCMC corner plots for all the cosmic string and superstring models considered. In each plot, the maximum-likelihood point is indicated by a star, while the corresponding values in the 1D marginals are shown with dotted red lines. The dashed black lines show the priors of each parameter.

\begin{figure*}
    \centering
    \includegraphics[width=0.8\linewidth]{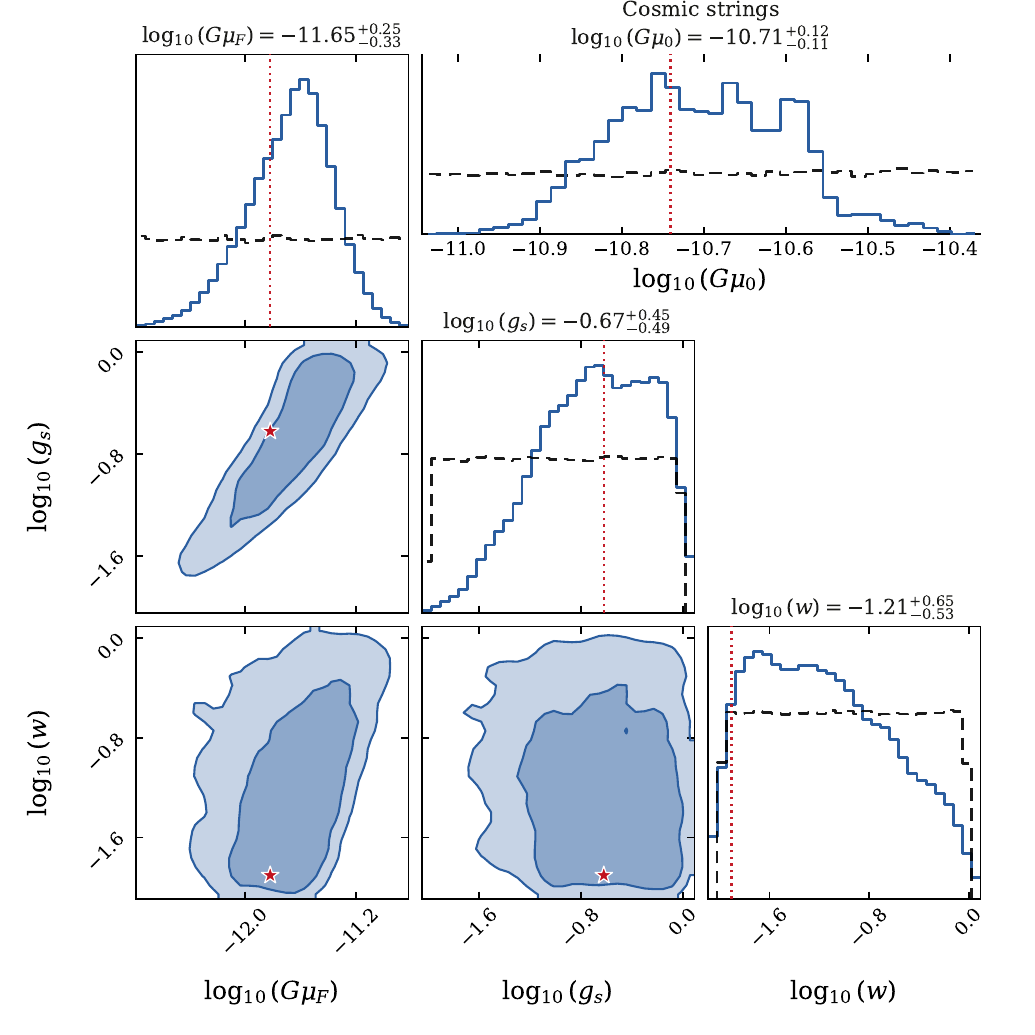}
    \caption{Posteriors for the cosmic string and superstring constant tension models fitted to NG15 and LVK O3 data. The top right panel shows the cosmic string single-parameter marginal, while the rest of the corner plot shows the superstring parameters. Quoted above each panel are the median and 68\% confidence level of each parameter.}
    \label{fig:const_corners}
\end{figure*}

\begin{figure*}
    \centering
    \includegraphics[width=0.8\textwidth]{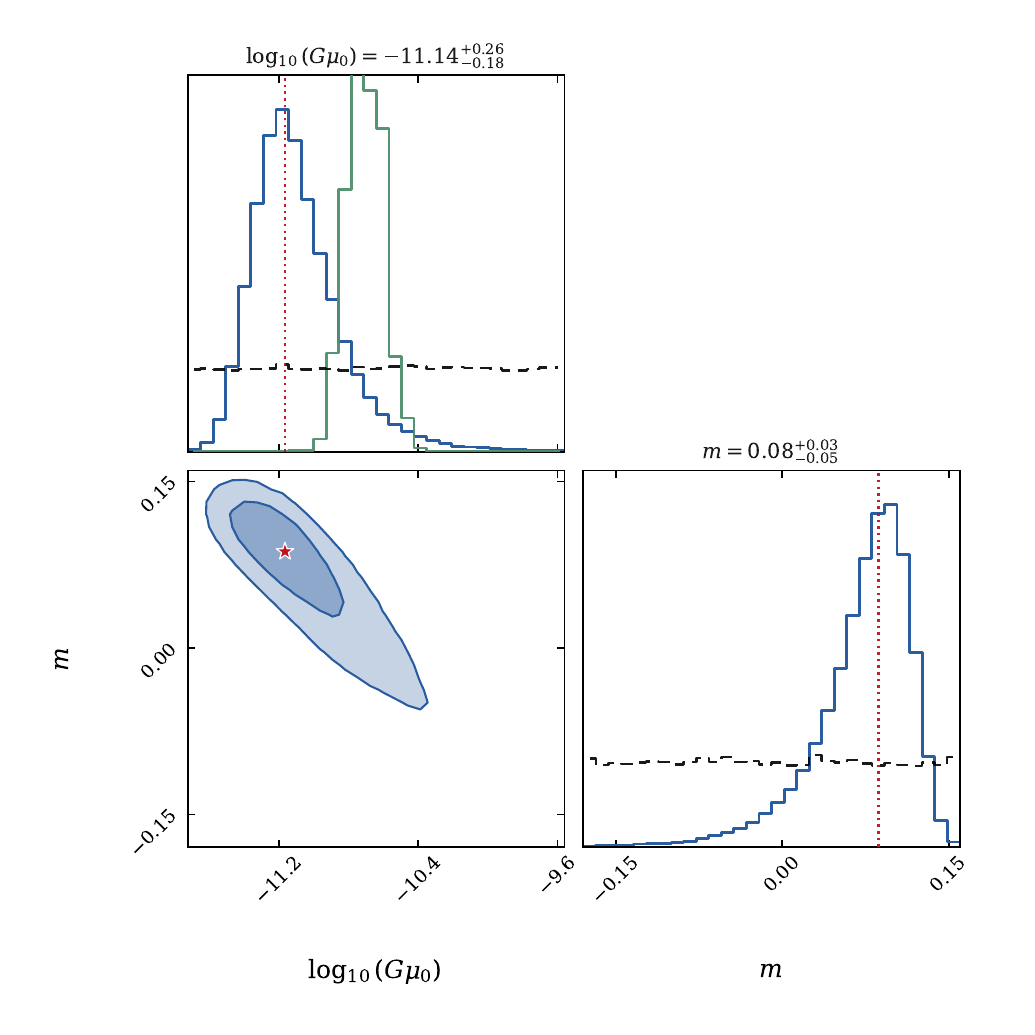}
    \caption{Posteriors for the cosmic string full time varying tension model fitted to NG15 and LVK O3 data. The green marginal shows the constant tension cosmic string case for comparison. Quoted above each panel are the median and 68\% confidence level of each parameter.}
    \label{fig:CSfull_corner}
\end{figure*}

\begin{figure*}
    \centering
    \includegraphics[width=0.8\textwidth]{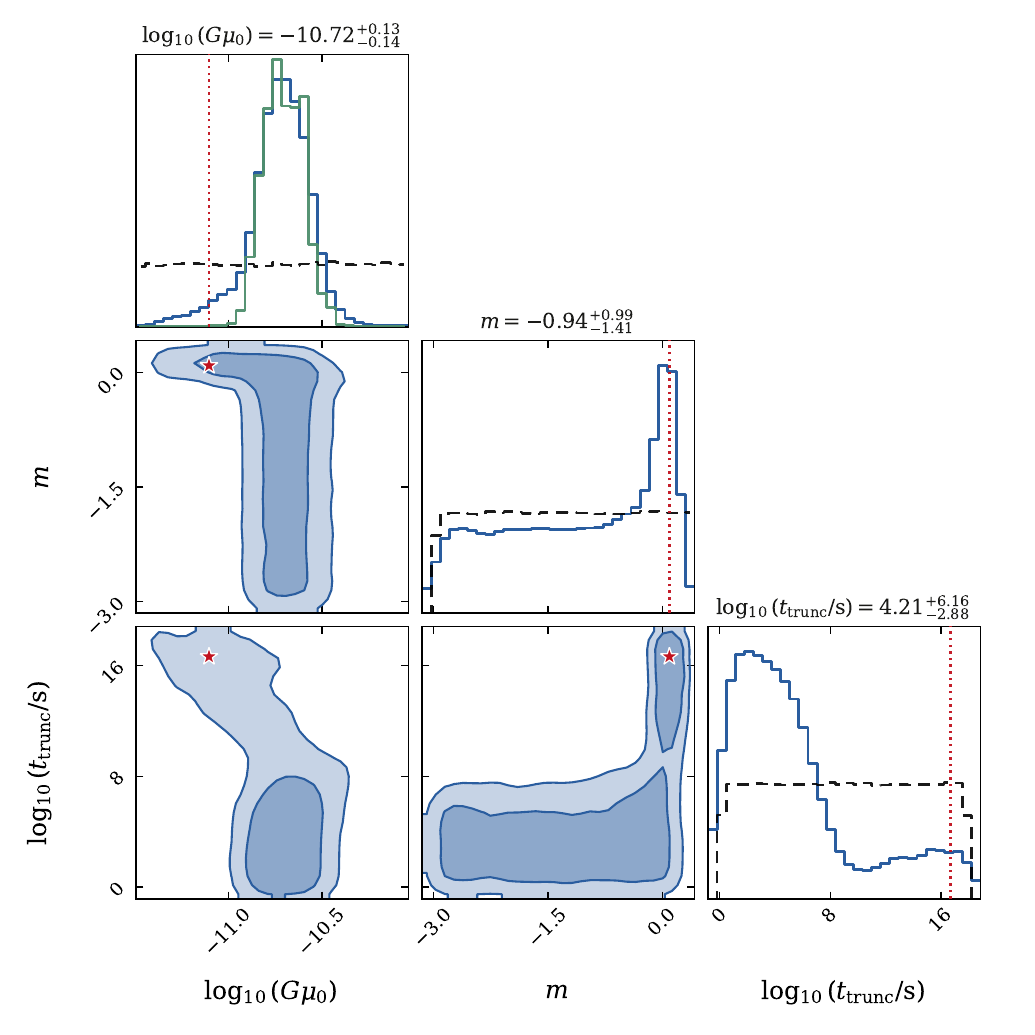}
    \caption{Posteriors for the cosmic string truncated time varying tension model fitted to NG15 and LVK O3 data. The green marginal shows the constant tension cosmic string case for comparison. Quoted above each panel are the median and 68\% confidence level of each parameter.}
    \label{fig:CStrunc_corner}
\end{figure*}

\begin{figure*}
    \centering
    \includegraphics[width=0.8\textwidth]{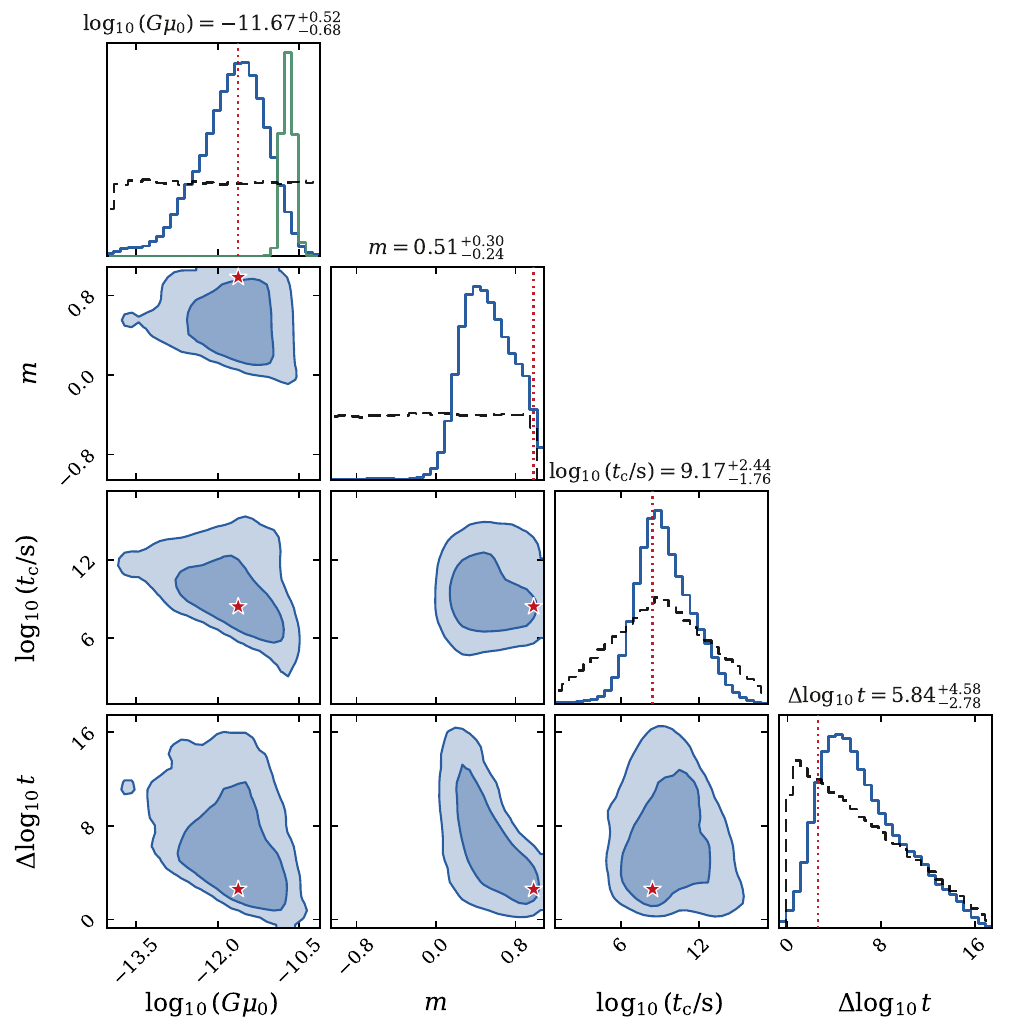}
    \caption{Posteriors for the cosmic string window time varying tension model fitted to NG15 and LVK O3 data. The green marginal shows the constant tension cosmic string case for comparison. Quoted above each panel are the median and 68\% confidence level of each parameter.}
    \label{fig:CSwindow_corner}
\end{figure*}

\begin{figure*}
    \centering
    \includegraphics[width=0.8\textwidth]{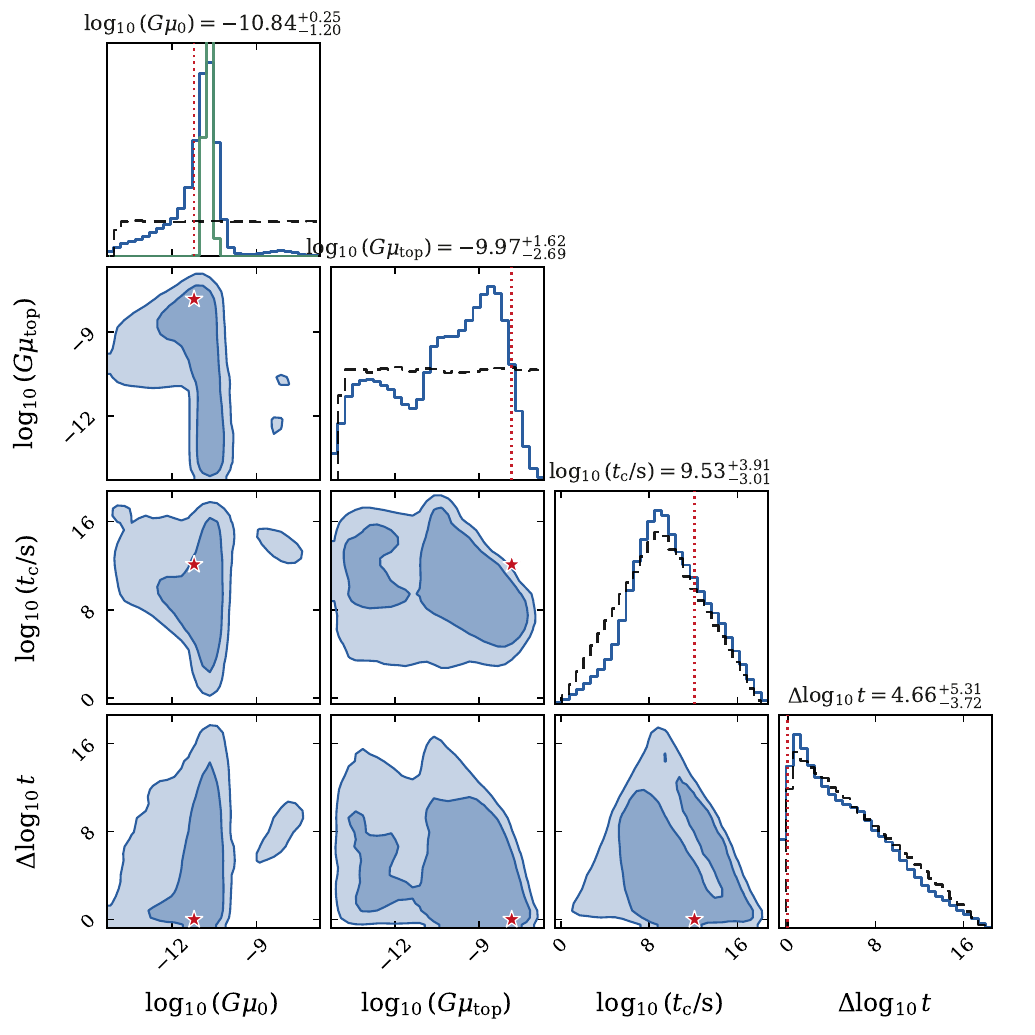}
    \caption{Posteriors for the cosmic string top-hat time varying tension model fitted to NG15 and LVK O3 data. The green marginal shows the constant tension cosmic string case for comparison. Quoted above each panel are the median and 68\% confidence level of each parameter.}
    \label{fig:CStophat_corner}
\end{figure*}

\begin{figure*}
    \centering
    \includegraphics[width=0.8\textwidth]{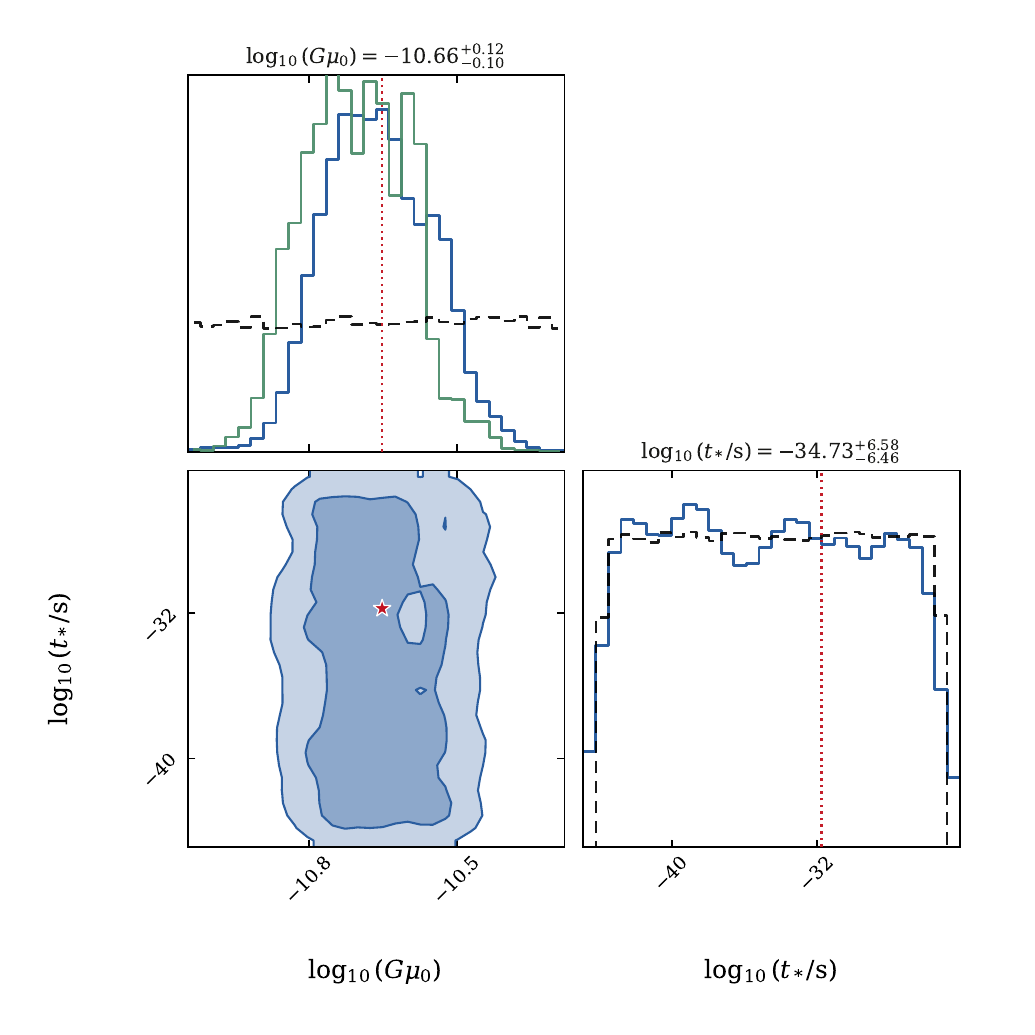}
    \caption{Posteriors for the cosmic string axion kination model fitted to NG15 and LVK O3 data. The green marginal shows the constant tension cosmic string case for comparison. Quoted above each panel are the median and 68\% confidence level of each parameter.}
    \label{fig:CSaxion_corner}
\end{figure*}

\begin{figure*}
    \centering
    \includegraphics[width=0.8\textwidth]{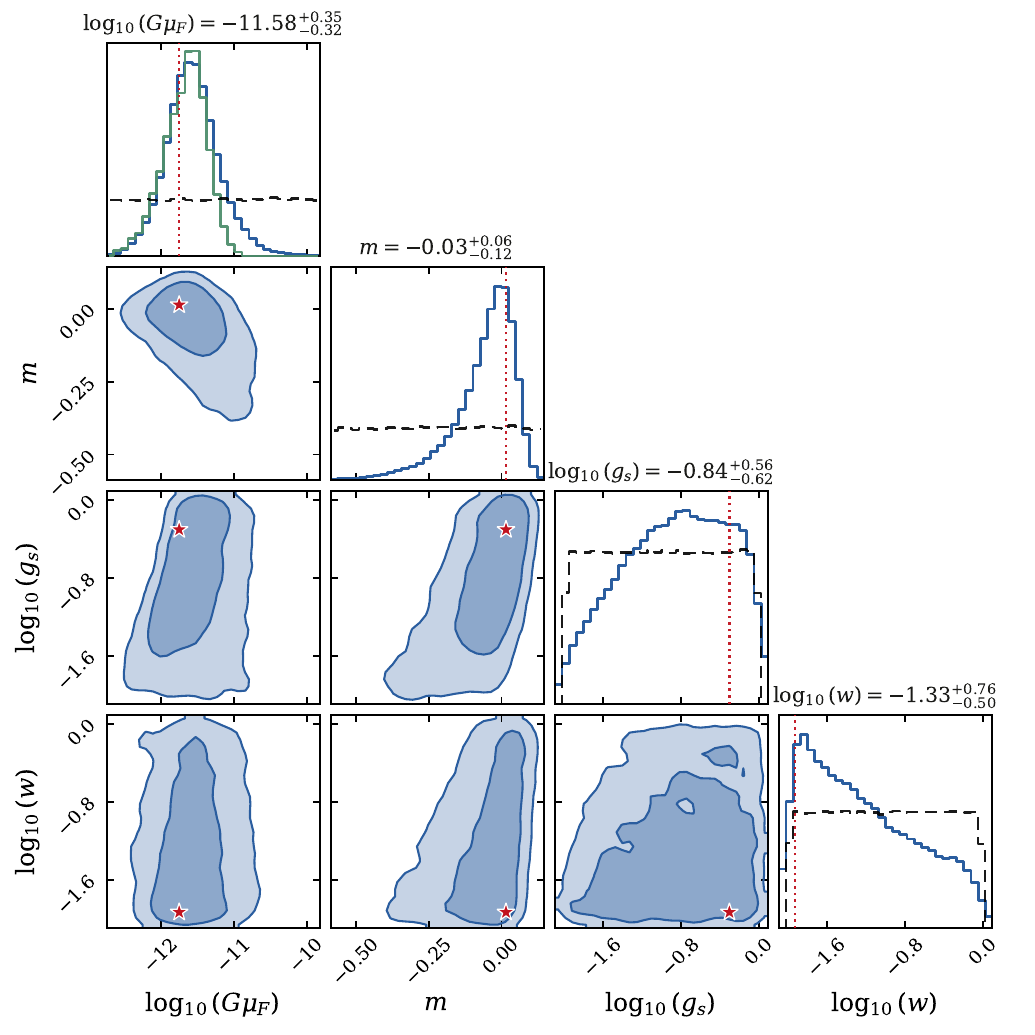}
    \caption{Posteriors for the cosmic superstring full time varying tension model fitted to NG15 and LVK O3 data. The green marginal shows the constant tension cosmic superstring case for comparison. Quoted above each panel are the median and 68\% confidence level of each parameter.}
    \label{fig:SSfull_corner}
\end{figure*}

\begin{figure*}
    \centering
    \includegraphics[width=0.8\textwidth]{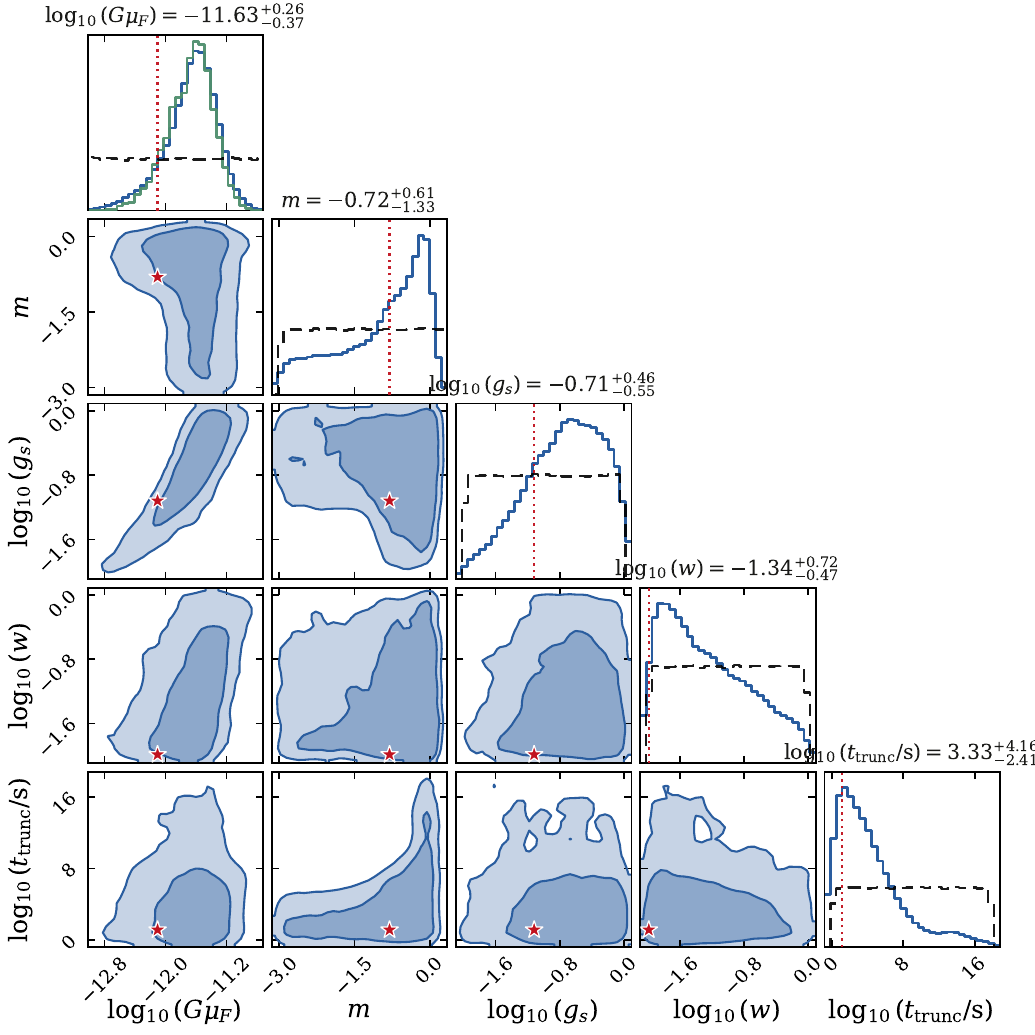}
    \caption{Posteriors for the cosmic superstring truncated time varying tension model fitted to NG15 and LVK O3 data. The green marginal shows the constant tension cosmic superstring case for comparison. Quoted above each panel are the median and 68\% confidence level of each parameter.}
    \label{fig:SStrunc_corner}
\end{figure*}

\begin{figure*}
    \centering
    \includegraphics[width=0.8\textwidth]{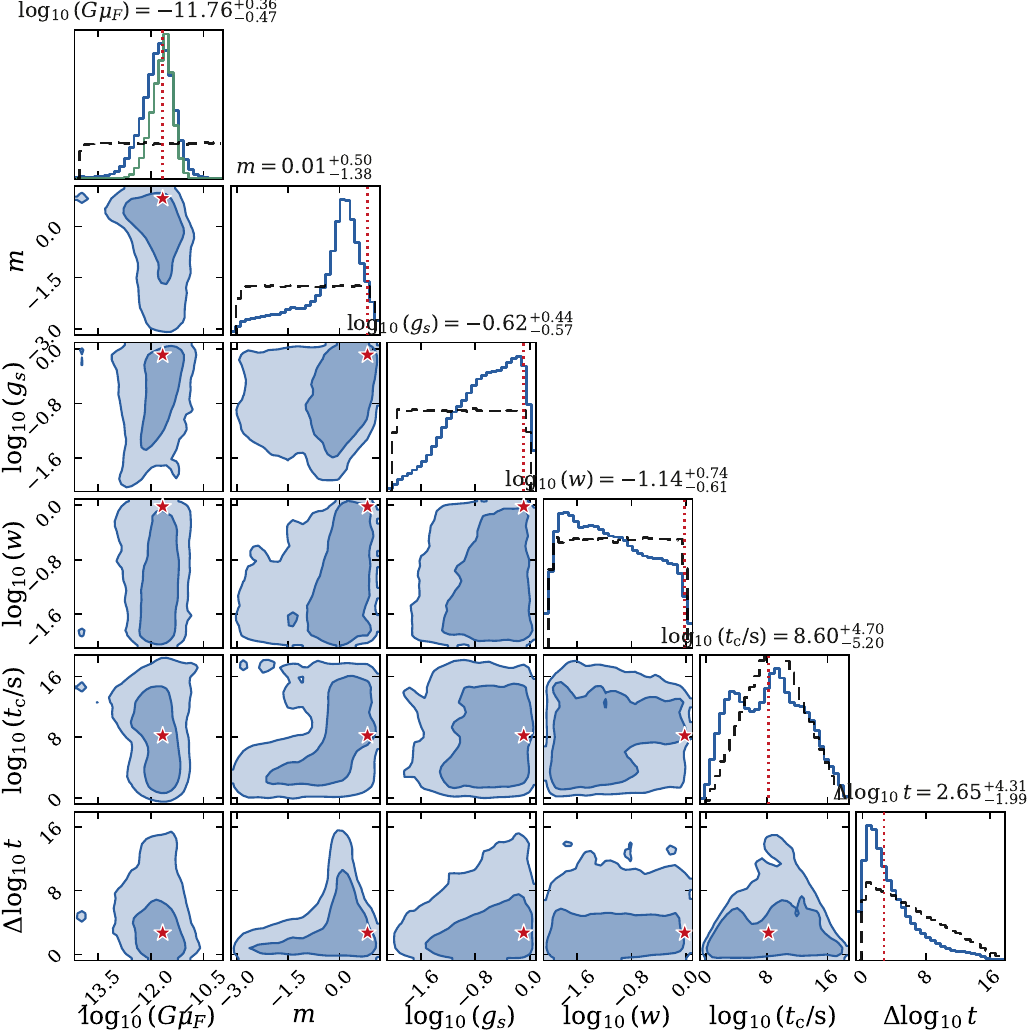}
    \caption{Posteriors for the cosmic superstring window time varying tension model fitted to NG15 and LVK O3 data. The green marginal shows the constant tension cosmic superstring case for comparison. Quoted above each panel are the median and 68\% confidence level of each parameter.}
    \label{fig:SSwindow_corner}
\end{figure*}

\begin{figure*}
    \centering
    \includegraphics[width=0.8\textwidth]{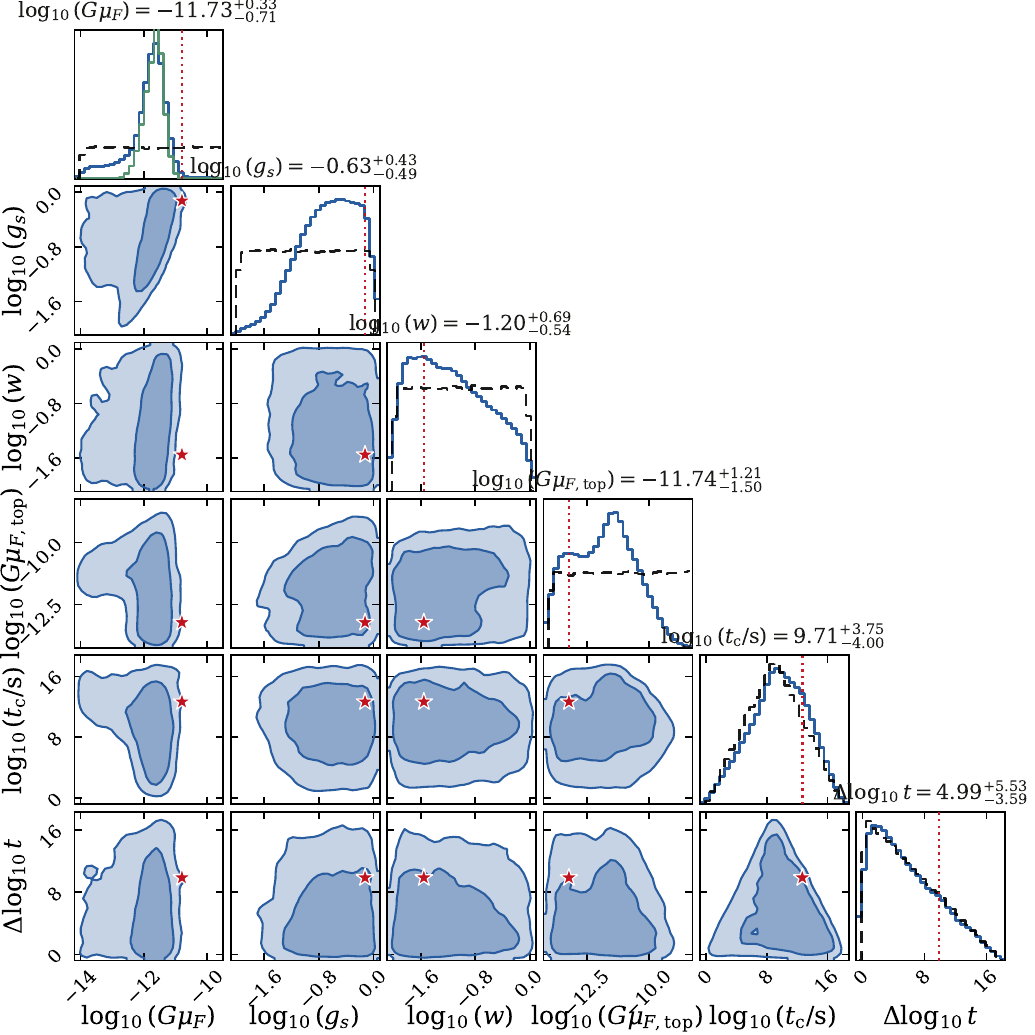}
    \caption{Posteriors for the cosmic superstring top-hat time varying tension model fitted to NG15 and LVK O3 data. The green marginal shows the constant tension cosmic superstring case for comparison. Quoted above each panel are the median and 68\% confidence level of each parameter.}
    \label{fig:SStophat_corner}
\end{figure*}

\begin{figure*}
    \centering
    \includegraphics[width=0.8\textwidth]{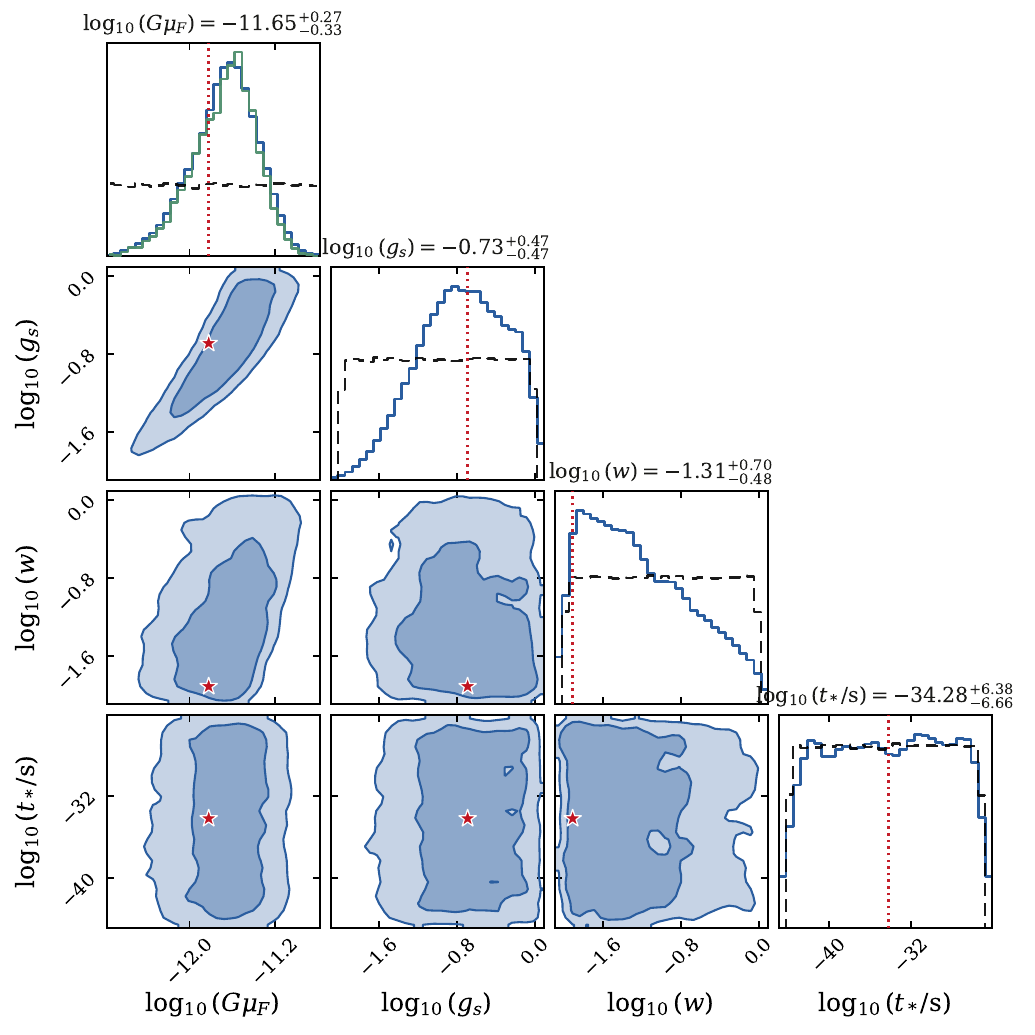}
    \caption{Posteriors for the cosmic superstring axion kination model fitted to NG15 and LVK O3 data. The green marginal shows the constant tension cosmic superstring case for comparison. Quoted above each panel are the median and 68\% confidence level of each parameter.}
    \label{fig:SSaxion_corner}
\end{figure*}

\bibliography{refs}
\end{document}